\documentclass[letterpaper, 10 pt, conference]{ieeeconf}  

\usepackage{cite}
\usepackage{amsmath,amssymb,amsfonts}
\usepackage{algorithmic}
\usepackage{graphicx}
\usepackage{textcomp}
\usepackage{bm}
\usepackage{hyperref}
\usepackage{url} 
\usepackage{subcaption}
\usepackage[table,xcdraw]{xcolor}
\usepackage{draftwatermark}
\SetWatermarkText{Accepted by American Control Conference (ACC) 2025}
\SetWatermarkScale{0.07}
\SetWatermarkLightness{0.6} 
\SetWatermarkAngle{0} 
\SetWatermarkHorCenter{300pt} 
\SetWatermarkVerCenter{10pt}

\IEEEoverridecommandlockouts                              

\title{\LARGE \bf
Robustness Enhancement for Multi-Quadrotor Centralized Transportation System via Online Tuning and Learning
}

\author{Tianhua Gao$^{1}$, Kohji Tomita$^{2}$ and Akiya Kamimura$^{3}$ 
\thanks{The authors are with the Industrial Cyber-Physical Systems Research Center, Department of Information Technology and Human Factors, National
Institute of Advanced Industrial Science and Technology (AIST), Japan (\{$^{1}$kou.tenka, $^{2}$k.tomita, $^{3}$kamimura.a\}@aist.go.jp). $^{1}$T. Gao is also with 
the Graduate School of
Systems and Information Engineering, University of Tsukuba, Japan ($^{1}$gao.tianhua@ieee.org, gao.tianhua@ieee.org, gao.tianhua.tkb\_gb@u.tsukuba.ac.jp). For simulation videos in Section \uppercase\expandafter{\romannumeral 5}, refer to \url{https://staff.aist.go.jp/kamimura.a/ACC/video.mp4}.
This paper has been accepted for publication in the Proceedings of the American Control Conference (ACC) 2025. Published in IEEE Xplore: \url{https://ieeexplore.ieee.org/document/11107693}.}%
}

\begin{document}

\maketitle
\thispagestyle{empty}
\pagestyle{empty}

\begin{abstract}

This paper introduces an  adaptive-neuro geometric control for a centralized multi-quadrotor cooperative transportation system, which enhances both adaptivity and disturbance rejection. Our strategy is to coactively tune the model parameters and learn the external disturbances in real-time. To realize this, we augmented the existing geometric control with multiple neural networks and adaptive laws, where the estimated model parameters and the weights of the neural networks are simultaneously tuned and adjusted online. The Lyapunov-based adaptation guarantees bounded estimation errors without requiring either pre-training or the persistent excitation (PE) condition.
The proposed control system has been proven to be stable in the sense of Lyapunov under certain preconditions, and its enhanced robustness under scenarios of disturbed environment and model-unmatched plant was demonstrated by numerical simulations.

\end{abstract}

\section{INTRODUCTION}

Cable-suspended payload transportation with multi-quadrotor systems has gained significant attention in the Unmanned Aerial Vehicles (UAVs) manipulation field\cite{2024 A Review of Real-Time Implementable Cooperative Aerial Manipulation Systems} due to its greater efficiency. Regarding the controller design in this area,  recent studies can be divided into two primary orientations: the decentralized methods (e.g.,\cite{2011 Cooperative manipulation and transportation with aerial robots}-\cite{2024 Decentralized adaptive controller for multi-drone cooperative transport with offset and moving center of gravity}) and the centralized methods (e.g., \cite{2013 Dynamics Control and Planning for Cooperative
Manipulation of Payloads Suspended by Cables
from Multiple Quadrotor Robots}-\cite{2024 Efficient Optimization-Based Cable Force Allocation for Geometric Control of a Multirotor Team Transporting a Payload}).

   Decentralized methods are more low-cost and easier to implement since they do not require state estimation of the complete dynamics of the cables and payload. Some studies\cite{2017 Collaborative transportation using MAVs via passive force control}, \cite{2022 Controller Design and Disturbance Rejection of
Multi-Quadcopters for Cable Suspended Payload
Transportation Using Virtual Structure} completely discard the feedback from the payload, adopting leader-follower schemes for configuration among multirotors. These algorithms are robust since they do not rely on the knowledge of the payload, but the payload pose remains uncontrolled during transportation. For this reason, in recent state-of-art, some researches have gradually focused on solving the payload pose manipulation problem.  In \cite{2020 Full-Pose Manipulation Control of a Cable-Suspended Load With Multiple UAVs Under Uncertainties}, \cite{2022 Indirect Force Control of a Cable-Suspended Aerial Multi-Robot Manipulator} the authors have achieved full-pose manipulation of a cable-suspended platform in a quasi-static condition. In \cite{2023 Force-Based Pose Regulation of a Cable-Suspended Load Using UAVs with Force Bias}, \cite{2023 Equilibria Stability and Sensitivity for the Aerial
Suspended Beam Robotic System Subject to
Parameter Uncertainty}, the pose regulation of a cable-suspended beam has been proposed utilizing the incomplete system dynamics.

By contrast, the centralized methods aim at  the dynamical tracking of the payload trajectory and orientation. These methods generally require the full dynamics of the system including cables, which was considered to be challenging. However, with the development of the sensor technology,  current studies (e.g., \cite{2024 Aerial Transportation of Cable-Suspended Loads With an Event Camera}) have been capable of  estimating the complete cable's state. Also, the feasibility of the vision-based centralized method has been validated by the real-world experiment \cite{2021 Cooperative Transportation of Cable Suspended Payloads With MAVs Using Monocular Vision and Inertial Sensing}. Furthermore, the collision problem, once considered a drawback of centralized methods, has been resolved by the optimization-based strategy in \cite{2024 Efficient Optimization-Based Cable Force Allocation for Geometric Control of a Multirotor Team Transporting a Payload}. Thus, the centralized methods are increasingly promising for the dynamic transportation of payloads with stable pose.

\begin{figure}[t]
      \centering
      \includegraphics[scale=0.2]{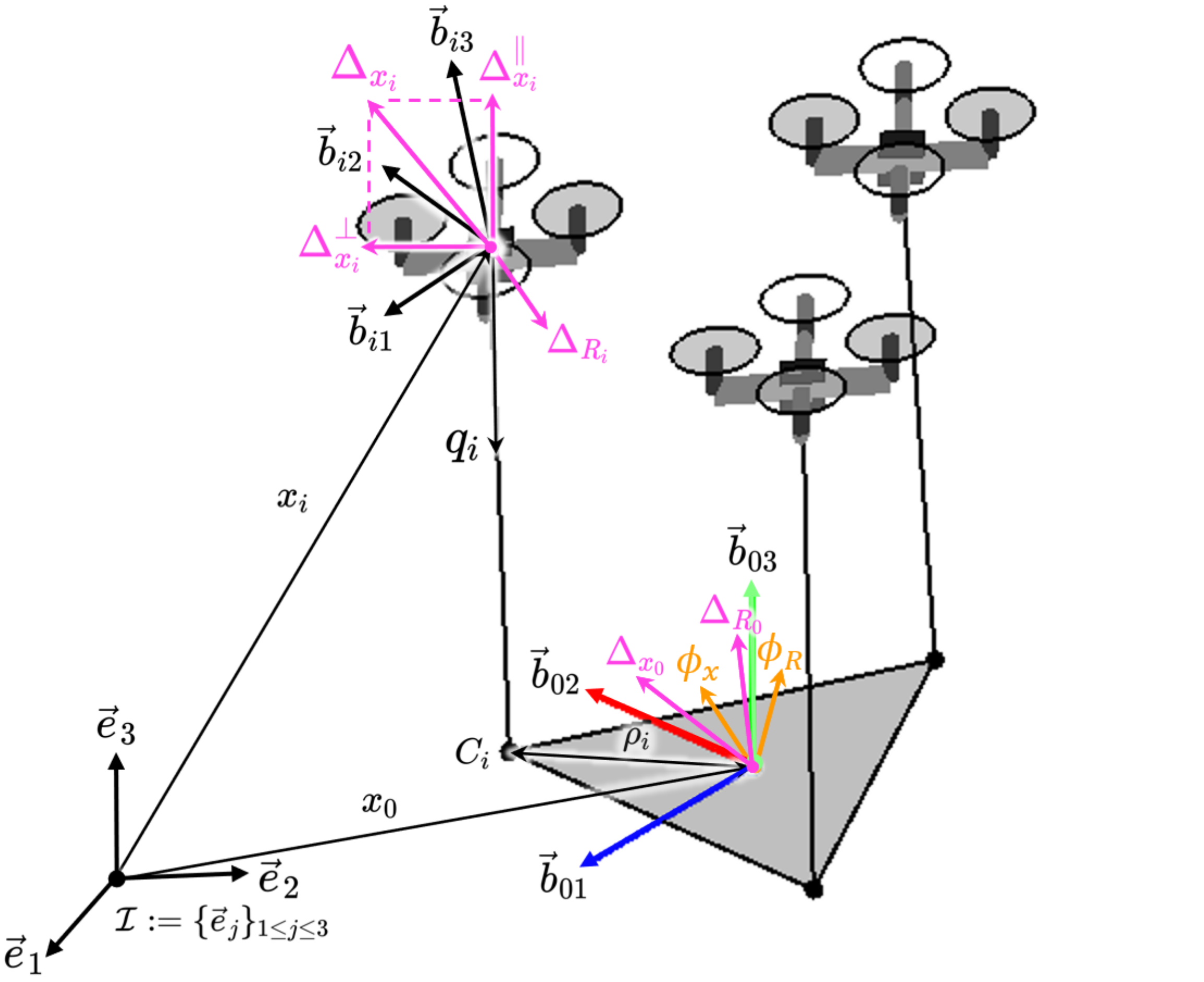}
      \caption{\footnotesize Dynamics model with  disturbances: $\Delta_{\bm{x}_{i}}$, $\Delta_{\bm{R}_i}$, $\Delta_{\bm{x}_{i}}^{\parallel}$, $\Delta_{\bm{x}_{i}}^{\bot}$, $\Delta_{\bm{x}_0}$, $\Delta_{\bm{R}_0}\in\mathbb{R}^3$ and augmented disturbance dynamics: $\bm{\phi}_{\bm{\textit{x}}}$, $\bm{\phi}_{\bm{\textit{R}}}\in \mathbb{R}^3$.  $\Delta_{\bm{x}_{i}}$  and $\Delta_{\bm{R}_i}$ denote the disturbance force  and  moment 
 exerted on $i^{th}$ quadrotor, respectively.  The disturbance force $\Delta_{\bm{x}_{i}}$ can be decomposed into parallel and normal components along the cable, denoted as $\Delta_{\bm{x}_{i}}^{\parallel}$ and  $\Delta_{\bm{x}_{i}}^{\bot}$.  Similarly, the payload experiences disturbance force  $\Delta_{\bm{x}_0}$ and moment $\Delta_{\bm{R}_0}$.  In $n$-quadrotor scenario,  an inertial frame $\mathcal{I}:=\{\bm{\vec{e}}_j\}_{1\leq j \leq 3}$, a payload body-fixed frame $\mathcal{B}_{\bm{0}}:=\{\bm{\vec{b}}_{\bm{0}1},\bm{\vec{b}}_{\bm{0}2},\bm{\vec{b}}_{\bm{0}3}\}$  and $n$ quadrotor body-fixed frames  $\mathcal{B}_{\bm{i}}:=\{\bm{\vec{b}}_{\bm{i}1},\bm{\vec{b}}_{\bm{i}2}, \bm{\vec{b}}_{\bm{i}3} \}, i\in[1, n]$ are defined for modeling. For notations of the symbols  in this figure, refer to TABLE \uppercase\expandafter{\romannumeral 1}.}
      \label{Gao1}
   \end{figure}  

In our research project, there is a demand to precisely transport a payload of variable and unknown weight in a wind-disturbed environment while maintaining a controlled pose. Therefore, centralized methods are closer to meeting our requirements.
However, the existing state-of-art \cite{2023 Composite Disturbance Rejection Control Strategy for Multi-Quadrotor Transportation System} has addressed only disturbance rejection but not parametric uncertainties. To fill this gap, we seek to further address both parametric uncertainties and disturbance rejection. Our contributions are summarized as follows:
\begin{itemize}
\item Proposed an adaptive-neuro geometric control using online tuning and learning to simultaneously estimate the parametric uncertainties and reject the external disturbances.
\item Proved the stability of the proposed control system in the absence of  parametric uncertainties and under conditions of static attitude tracking. Analyzed the robust stability in the presence of parametric uncertainties and disturbances under dynamic tracking conditions.
\item Demonstrated the enhanced robustness through simulations.
\end{itemize}

This paper is organized as follows. Section \uppercase\expandafter{\romannumeral 2} describes the system dynamics. Section \uppercase\expandafter{\romannumeral 3} introduces the control strategy. Section \uppercase\expandafter{\romannumeral 4} analyzes the stability of the proposed control system. The enhanced robustness is demonstrated by simulations in Section \uppercase\expandafter{\romannumeral 5}. Finally, Section \uppercase\expandafter{\romannumeral 6} concludes the paper and discusses future work.

\section{Dynamics with Augmented Disturbance }

This section introduces the disturbance-augmented full dynamics for the multi-quadrotor cooperative transportation system. 
In recent works, the centralized geometric control strategy proposed by \cite{2018 Geometric Control of Quadrotor UAVs Transporting
a Cable-Suspended Rigid Body} has been widely studied in state of the art and its feasibility has been verified by experiments \cite{2021 Cooperative Transportation of Cable Suspended Payloads With MAVs Using Monocular Vision and Inertial Sensing}, \cite{2023 Nonlinear Model Predictive Control for Cooperative Transportation and Manipulation of Cable Suspended Payloads with Multiple Quadrotors} and open-source simulations\cite{2024 RotorTM: A Flexible Simulator for Aerial Transportation and Manipulation}.
In this study, we further augment the dynamics in \cite{2018 Geometric Control of Quadrotor UAVs Transporting
a Cable-Suspended Rigid Body} with unknown translational and rotational disturbance dynamics terms  $\bm{\phi}_{\bm{\textit{x}}}$, $\bm{\phi}_{\bm{\textit{R}}}\in \mathbb{R}^3$, and then reformulate the full dynamics as follows:
\begin{equation}
{\footnotesize
  \begin{cases}
  \bm{\dot{\Omega}}_i=\bm{J}_i^{-1}(\mathbf{M}_i-\bm{\Omega}_i\times\bm{J}_i\bm{\Omega}_i+\Delta_{\bm{R}_i}),\\[1pt]
 \bm{\dot{R}}_i=\bm{R}_i[\bm{\Omega}_i]_{\times},\\[1pt]

   \bm{\ddot{q}}_i=\frac{1}{m_il_i}[\bm{q}_i]_{\times}^2(\mathbf{u}_i+\Delta_{\bm{x}_{i}}-m_i\bm{a}_i)-\lVert \bm{\dot{q}}_i \rVert^2_2\bm{q}_i,
   \\[3pt]
\bm{\dot{\omega}}_i=\frac{1}{l_i}[\bm{q}_i]_{\times}\bm{a}_i-\frac{1}{m_il_i}[\bm{q}_i]_{\times}(\mathbf{u}_i^{\bot}+\Delta_{\bm{x}_{i}}^{\bot}),\\[3pt]
\bm{\ddot{x}}_0=\frac{1}{m_0}\left(\bm{\mathrm{F}_d}+\Delta_{\bm{x}_0}+\bm{\displaystyle \sum}_{i=1}^n\Delta_{\bm{x}_{i}}^{\parallel}\right)-g\bm{\vec{e}}_3+\bm{Y_{\textit{x}}}+\bm{\phi}_{\bm{\textit{x}}},
\\[5pt]
\begin{aligned}
    \bm{\dot{\Omega}}_0=&\bm{J}_0^{-1}\left(\bm{\mathrm{M}_d}-[\bm{\Omega}_0]_{\times}\bm{J}_0\bm{\Omega}_0+\Delta_{\bm{R}_0}+\bm{\displaystyle \sum}_{i=1}^n[\bm{\rho}_i]_{\times}\bm{R}_0^{\top}\Delta_{\bm{x}_{i}}^{\parallel}\right)\\
   &+\bm{Y_{\textit{R}}}+\bm{\phi}_{\bm{\textit{R}}},
\end{aligned}
 \\[4pt]
  \bm{\dot{R}}_0=\bm{R}_0[\bm{\Omega}_0]_{\times},
    \end{cases}
    \label{disturbance-augmented dinamics}
    }
\end{equation}
where $\Delta_{\bm{x}_{i}}$, $\Delta_{\bm{R}_i}$,  $\Delta_{\bm{x}_{i}}^{\parallel}$, $\Delta_{\bm{x}_{i}}^{\bot}$,  $\Delta_{\bm{x}_0}$, $\Delta_{\bm{R}_0}\in\mathbb{R}^3$ are bounded disturbances, as illustrated in Fig. \ref{Gao1}; $\bm{\mathrm{F}_d}$ and $\bm{\mathrm{M}_d}\in\mathbb{R}^3$ are the desired resultant control force and moment acting on the payload, which will be designed as first-level control signals in Section \ref{Adaptive-Neuro geometric control}; $\bm{a}_i$ is the acceleration of $i^{th}$ connection point $C_i$;  $\mathbf{u}_i^{\parallel}=(\bm{q}_{i}\otimes\bm{q}_i)\mathbf{u}_i$ and $\mathbf{u}_i^{\bot}=(\mathbf{I}^{3\times3}-\bm{q}_{i}\otimes\bm{q}_i)\mathbf{u}_i$ are the parallel and normal components of $\mathbf{u}_i$ with respect to vector $\bm{q}_i$;  $\bm{Y_{\textit{x}}}$ and $\bm{Y_{\textit{R}}}\in\mathbb{R}^3$ are errors caused by tracking deviations of cables, respectively:
\begin{equation}
{\footnotesize
    \begin{aligned}
&\bm{a}_i=\bm{\ddot{x}}_0+g\bm{\vec{e}}_3+\bm{R}_0[\bm{\Omega}_0]_{\times}^2\bm{\rho}_{i}-\bm{R}_0[\bm{\rho}_i]_{\times}\bm{\dot{\Omega}}_0,\\
     &\mathbf{u}_i^{\parallel} =\bm{\mathrm{\mu}}_i+m_il_i\lVert \bm{\omega}_i \rVert^2_2\bm{q}_i-m_i(\bm{q}_{i}\otimes \bm{q}_i) \bm{a}_i,\\
&\mathbf{u}_i=\mathbf{u}_i^{\parallel}+\mathbf{u}_i^{\bot},\\
      &\bm{Y_{\textit{x}}}=\frac{1}{m_0}\bm{\sum}_{i=1}^n\left(\bm{\mathrm{\mu}}_{i}-\bm{\mathrm{\mu}}_{i_{\bm{d}}}\right), \\
&\bm{Y_{\textit{R}}}=\bm{J}_0^{-1}\bm{\sum}_{i=1}^n[\bm{\rho}_i]_{\times}\bm{R}_0^{\top}\left(\bm{\mathrm{\mu}}_{i}-\bm{\mathrm{\mu}}_{i_{\bm{d}}}\right),
  \label{Acceleration and Parallel Component}
\end{aligned}
}
\end{equation}
in which the internal tension along the $i^{th}$  cable $\bm{\mathrm{\mu}}_{i}\in \mathbb{R}^3=(\bm{q}_{i}\otimes\bm{q}_i)\bm{\mathrm{\mu}}_{i_{\bm{d}}}$. Its  desired value $\bm{\mathrm{\mu}}_{i_{\bm{d}}}\in\mathbb{R}^3$ satisfies $\bm{\sum}_{i=1}^n\bm{\mathrm{\mu}}_{i_{\bm{d}}}=\bm{\mathrm{F}_d}$, $ \bm{\sum}_{i=1}^n[\bm{\rho}_i]_{\times}\bm{R}_0^{\top}\bm{\mathrm{\mu}}_{i_{\bm{d}}}=\bm{\mathrm{M}_d}$ and is solved by the following minimum-norm solution: 
\begin{equation}
    {\footnotesize
    \begin{aligned}
 &{\renewcommand{\arraystretch}{0.1}\begin{bmatrix}\bm{\mathrm{\mu}}_{1_{\bm{d}}} \\\cdot\\\cdot\\ \bm{\mathrm{\mu}}_{n_{\bm{d}}}\\\end{bmatrix}}=\bm{\mathrm{diag}}{\renewcommand{\arraycolsep}{0.4pt}\begin{bmatrix}
         \bm{R_0} &\cdot&\cdot&\bm{R_0}
     \end{bmatrix}}{\renewcommand{\arraycolsep}{0.4pt}\begin{bmatrix}
\mathbf{I}^{3\times3}&\cdot&\cdot&\mathbf{I}^{3\times3}\\
        [\bm{\rho}_{1}]_{\times} &\cdot&\cdot& [\bm{\rho}_{n}]_{\times} 
    \end{bmatrix}}^\dagger\begin{bmatrix} \bm{R}_0^{\top}\bm{\mathrm{F}_d} \\ \bm{\mathrm{M}_d}\end{bmatrix}.
\end{aligned}
    }
\end{equation}
Through the foregoing  rearrangement, $\{\bm{\mathrm{F}_d}, \bm{\mathrm{M}_d}\}$ are explicitly  included in the system dynamics, instead of $\bm{\mathrm{\mu}}_{i}$ or $\mathbf{u}_i^{\parallel}$. For symbols not elaborated, refer to TABLE {\uppercase\expandafter{\romannumeral 1} and Fig. \ref{Gao1}.
\begin{table}[thbp]
\label{table1}
  \centering
  \caption{Symbol References and Notations}
  \begin{tabular}{>{\columncolor{gray!10}}c|l}
    \hline
    \hline
      $C_i\in{E}^3$&Connection point between payload and $i^{th}$ cable\\
    $g\in\mathbb{R}$ &Gravitational acceleration\\
    $l_i\in\mathbb{R}$ &Length of $i^{th}$ cable\\
    $m_0,m_i\in\mathbb{R}$ & Mass of payload and $i^{th}$ quadrotor\\
    $\bm{x}_0,\bm{x}_i\in\mathbb{R}^3$ & Position of payload, $i^{th}$ quadrotor in $\mathcal{I}$\\
    $\bm{\dot{x}}_0,\bm{\dot{x}}_i\in\mathbb{R}^3$ & Linear velocity of payload, $i^{th}$ quadrotor in $\mathcal{I}$ \\
    $\bm{\ddot{x}}_0,\bm{\ddot{x}}_i\in\mathbb{R}^3$ & Linear acceleration of payload, $i^{th}$ quadrotor in $\mathcal{I}$ \\
    $\bm{\Omega}_0,\bm{\dot{\Omega}}_0\in\mathbb{R}^3$ & Angular velocity, acceleration of payload in $\mathcal{I}$\\
    $\bm{\Omega}_i\in\mathbb{R}^3$& Angular velocity of $i^{th}$ quadrotor in $\mathcal{I}$ \\
     $\bm{\omega}_i\in\mathbb{R}^3$ & Angular velocity of $i^{th}$ cable in $\mathcal{B}_i$ \\
    $\bm{\rho}_i\in\mathbb{R}^3$ & Position of  $C_i$ in $\mathcal{B}_0$\\
    $\bm{J}_0,\bm{J}_i\in\mathbb{R}^{3\times3}$ & Inertia tensor of payload, $i^{th}$ quadrotor\\
    $\bm{R}_0\in \mathbf{SO}(3)$ & Rotation Matrix of $\mathcal{B}_0$ relative to $\mathcal{I}$\\
    $\bm{R}_{i}\in \mathbf{SO}(3)$ & Rotation Matrix of $\mathcal{B}_i$ relative to $\mathcal{I}$\\
    $\bm{q}_i\in \mathbf{S}^2$ &Unit vector from $i^{th}$ quadrotor to $C_i$ in $\mathcal{I}$\\
     $\mathbf{u}_i\in\mathbb{R}^3$ & Control force at $i^{th}$ quadrotor \\
    $\mathbf{M}_i\in\mathbb{R}^3$ & Control moment at $i^{th}$ quadrotor \\
    \hline
    \hline
     \multicolumn{2}{{p{225pt}}}{The notations used in this paper are listed as follows:
     
     Symbol $\otimes$ denotes the tensor product. 

     Superscript $\bullet^{\dagger}$ denotes the  pseudoinverse of a matrix.
     
     Symbols with subscript $\bullet_i$ relate to the $i^{th}$ quadrotor for $i\in[1, n]$.

     Symbols with subscript $\bullet_0$ relate to the payload.

     Symbols with $\bullet^{\parallel}_i$ represent quantities parallel to $\bm{q}_i$.
     
     Symbols with $\bullet^{\bot}_i$ represent quantities perpendicular to $\bm{q}_i$.

     Symbols with $\tiny\text{max}$ and $\tiny \text{min}$ denote the maximum and minimum values.
     
     Skew-symmetric map $[\,\bullet\,]_{\times}:\mathbb{R}^3\to\mathfrak{so}(3)$ is defined by the condition that $[\mathfrak{a}]_{\times}\mathfrak{b}=\mathfrak{a}\times\mathfrak{b}, \forall\mathfrak{a},\mathfrak{b}\in\mathbb{R}^3$.

       Element extraction map $\bullet^{[\cdot]}: (\mathbb{R}^3\cup\mathbb{R}^{3\times 3})\times\mathbb{N}\to\mathbb{R}$ extracts the $\cdot^{th}$ element from either a vector or the main diagonal of a matrix.
       
Vee map $\bullet^\vee: \mathfrak{so}(3)\to\mathbb{R}^3$ is defined as the inverse of skew-symmetric map.}
  \end{tabular}
\end{table}

\section{Adaptive-Neuro geometric control with Multiple Neural Networks}
\label{Adaptive-Neuro geometric control}
The existing geometric control \cite{2018 Geometric Control of Quadrotor UAVs Transporting
a Cable-Suspended Rigid Body} adopts a  Proportional-Integral-Differential (PID)-driven multi-level control flow:  $\{\bm{\mathrm{F}_d}, \bm{\mathrm{M}_d}\}\!\!\to\!\!\{\bm{\mathrm{\mu}}_{i_{\bm{d}}}\}_{1\leq i\leq n}\!\!\to\!\!\{\bm{\mathrm{\mu}}_{i}\}_{1\leq i\leq n}\!\!\to\!\!\{\mathbf{u}_i^{\parallel}, \mathbf{u}_i^{\bot}\}_{1\leq i\leq n}\!\to\!\{\bm{f}_i,\mathbf{M}_i\}_{1\leq i\leq n} $, where $\{\bm{\mathrm{F}_d}, \bm{\mathrm{M}_d}\}$ are the first-level PID control signals  for the desired payload state and $\{\bm{f}_i,\mathbf{M}_i\}_{1\leq i\leq n}$ are the final-level  control input thrusts and moments for the quadrotors. For the design of $\{\mathbf{u}_i^{\bot}, \bm{f}_i,\mathbf{M}_i\}_{1\leq i\leq n}$, refer to \cite{2018 Geometric Control of Quadrotor UAVs Transporting
a Cable-Suspended Rigid Body} and \cite{2021 Cooperative Transportation of Cable Suspended Payloads With MAVs Using Monocular Vision and Inertial Sensing}. 

Our aim is to enhance the robustness of the payload tracking by improving the first-level control signals $\{\bm{\mathrm{F}_d}, \bm{\mathrm{M}_d}\}$ without modifying $\{\bm{\mathrm{\mu}}_{i},\mathbf{u}_i^{\parallel}, \mathbf{u}_i^{\bot}, \bm{f}_i,\mathbf{M}_i\}_{1\leq i\leq n}$. In recent studies \cite{2021 Geometric Adaptive Control With Neural Networks
for a Quadrotor in Wind Fields} and \cite{2023 Quadrotor Neural Network Adaptive Control: Design
and Experimental Validation}, the multilayer Neural Networks (NNs) have been employed for disturbance rejection of quadrotors.  To reduce the burden of NNs and address parametric uncertainties, we further introduce Adaptive Laws (AL) to deploy an adaptive-neuro control strategy with $\mathrm{AL}\times(\mathrm{PD}-\mathrm{NNs})$ structure, as illustrated in Fig.~\ref{Gao2}.  Then, we append integral compensations to give the enhanced  $\{\bm{\mathrm{F}_d}, \bm{\mathrm{M}_d}\}$ in Eqs.~\eqref{Payload Translational Control} and \eqref{Payload Rotational Control}. 

\textbf{\textit{Notation 1:}} In the following text, symbols with superscripts $\bullet^{[j]}$, $1\leq j \leq 3$  and $\bullet^{[k]}$, $1\leq k \leq l$ denote the element extraction map as noted in TABLE \uppercase\expandafter{\romannumeral 1}. The position, attitude and angular velocity tracking errors of payload $\bm{e}_{\bm{x}_0}$, $\bm{e}_{\bm{R}_0}$, $\bm{e}_{\bm{\Omega}_0}\in\mathbb{R}^3$ are defined as in \cite{2013 Geometric nonlinear PID control of
a quadrotor UAV on SE(3)} and summarized here:
\begin{equation}
{\footnotesize
\begin{aligned}
\bm{e}_{\bm{x}_0}:=&\bm{x}_0-\bm{x}_{0_{\bm{d}}}, \\
     \bm{e}_{\bm{R}_0}:=&\frac{1}{2}(\bm{R}_{0_{\bm{d}}}^{\top}\bm{R}_0-\bm{R}_0^{\top}\bm{R}_{0_{\bm{d}}})^\vee , \bm{e}_{\bm{\Omega}_0}:=\bm{\Omega}_0-\bm{R}_0^{\top}\bm{R}_{0_{\bm{d}}}\bm{\Omega}_{0_{\bm{d}}},\notag
    \label{errors}
\end{aligned}
}
\end{equation}
where subscript $\bullet_d$ denotes the desired value.
\subsection{Neural Networks Formulation}
To approximate the unknown augmented disturbance dynamics terms $\bm{\phi}_{\bm{\textit{x}}}$ and $\bm{\phi}_{\bm{\textit{R}}}$ in Eq.~\eqref{disturbance-augmented dinamics}, multiple Radial Basis Function (RBF) neural networks with 2 inputs-$l$ hidden layer neurons-1 output (2-$l$-1) structure are deployed as follows:
\begin{equation}
{\footnotesize
\begin{aligned}
\bm{\phi}_{\circ}^{[j]}=\bm{\mathcal{W}}_{\circ j}^{\top}\bm{\hbar}(\textbf{x}_{\circ j})+\epsilon_{\circ j},
\label{Phi}
\end{aligned}
}
\end{equation}
where subscript $\bullet_{\circ\in\{\bm{\textit{x}}, \bm{\textit{R}}\}}$, refers to the symbols relating to translational and rotational dynamics. 
 $\textbf{x}_{\circ j}\in \mathbb{R}^2$ is the input vector of $j^{th}$ neural network, and  $\bm{\mathcal{W}}_{\circ j}\in\mathbb{R}^{l}$, $\epsilon_{\circ j}\in\mathbb{R}$, $\bm{\hbar}(\textbf{x}_{\circ j})\in\mathbb{R}^l$  are corresponding weights vector, bounded intrinsic approximation error and  Gaussian activation function, respectively.

The output of $k^{th}$ hidden layer neurons is given as follows:
\begin{equation}
{\footnotesize
\begin{aligned}
&\bm{\hbar}^{[k]}(\textbf{x}_{\circ j}):=\mathrm{exp}\left(-\frac{\lVert\textbf{x}_{\circ j}-\textbf{c}_{k}\rVert^2}{2b^{2}_k}\right),
\end{aligned}
\label{Gaussian activation function}
}
\end{equation}
where $\textbf{c}_{k}\in\mathbb{R}^2$ is the center vector of $k^{th}$ neurons and $b_k\in\mathbb{R}$ is the width of $k^{th}$ Gaussian function, $1\leq k \leq l$. 

To approximate Eq.~\eqref{Phi}, the estimated disturbance dynamics $\bm{\bar{\phi}}_{\circ}^{[j]}$ is given by the following neural network with time-varying estimated weight $\bm{\bar{\mathcal{W}}}_{\circ j}\in\mathbb{R}^{l}$:
\begin{equation}
{\footnotesize
\begin{aligned}
\bm{\bar{\phi}}_{\circ}^{[j]}:=\bm{\bar{\mathcal{W}}}_{\circ j}^{\top}\bm{\hbar}(\textbf{x}_{\circ j}),\\[-5pt]
\end{aligned}
\label{Phi_hat}
}
\end{equation}
where input  $\textbf{x}_{\bm{\textit{x}} j}:=\bm{\mathcal{E}}_{\bm{\textit{x}} j}:= \left(\bm{e}_{\bm{x}_0}^{[j]}, \bm{\dot{e}}_{\bm{x}_0}^{[j]}\right)^{\top}\in\mathbb{R}^2$ takes  the translational error vector along $\bm{\vec{e}}_j$-axis, and input $\textbf{x}_{\bm{\textit{R}} j}:= \left(\bm{e}_{\bm{R}_0}^{[j]}, \bm{e}_{\bm{\Omega}_0}^{[j]}\right)^{\top}$ takes the rotational error vector along $\bm{\vec{b}}_{\bm{0}j}$-axis. Both $\textbf{x}_{\bm{\textit{x}} j}$ and $\textbf{x}_{\bm{\textit{R}} j}$ are bounded by a compact set, satisfying the conditions required by the universal approximation theorem \cite{1989 Multilayer feedforward networks are universal approximators}. 

   \subsection{ First-level Control $\{\mathrm{F}_d, \mathrm{M}_d\}$}
   The adaptive-neuro control strategy for translational and rotational tracking, as illustrated in  Fig.~\ref{Gao2}, is designed in the form of the $j^{th}$ element as follows:
   \begin{equation}
{\footnotesize
\begin{aligned}
\bm{\mathcal{U}_{\bm{\textit{x}}}}^{[j]}:=\bm{\bar{m}}_0^{[j]}
\left(\bm{-\mathcal{K}}_{\bm{\textit{x}} j}^{\top}\bm{\mathcal{E}}_{\bm{\textit{x}} j}+\bm{\ddot{x}}_{0_{\bm{d}}}^{[j]}+g\bm{\delta}_{j3} -\bm{\bar{\phi}}_{\bm{\textit{x}}}^{[j]}\right),
\end{aligned}
}
\label{Translational Adaptive-Neuro Control}
\end{equation}
\begin{equation}
{\footnotesize
\begin{aligned}
\bm{\mathcal{U}_{\bm{\textit{R}}}}^{[j]}:=&\bm{\bar{J}}_0^{[j]}
\bigg{\{}-k_{R_0}\bm{e}^{[j]}_{\bm{R}_0}-k_{\Omega_0}\bm{e}^{[j]}_{\bm{\Omega}
_0}-\left([\bm{\Omega}_0]_{\times}\bm{R}_0^{\top}\bm{R}_{0_{\bm{d}}}\bm{\Omega}_{0_{\bm{d}}}\right)^{[j]}\\
&+\left(\bm{R}_0^{\top}\bm{R}_{0_{\bm{d}}}\bm{\dot{\Omega}}_{0_{\bm{d}}}\right)^{[j]}
-\bm{\bar{\phi}}_{\bm{\textit{R}}}^{[j]}\bigg{\}},
\end{aligned}
}
\label{Rotational Adaptive-Neuro Control}
\end{equation}
where  $\bm{\bar{m}}_0^{[j]}$ and $\bm{\bar{J}}_0^{[j]}$ are the $j^{th}$ element of the estimated payload mass vector $\bm{\bar{m}}_0\in \mathbb{R}^{3\times1}$ and the $j^{th}$  diagonal element of the estimated payload inertia tensor $\bm{\bar{J}}_0\in\mathbb{R}^{3\times3}$, respectively, tuned according to the following adaptive laws:
\begin{figure}[tbp]
      \centering
      \includegraphics[scale=0.148]{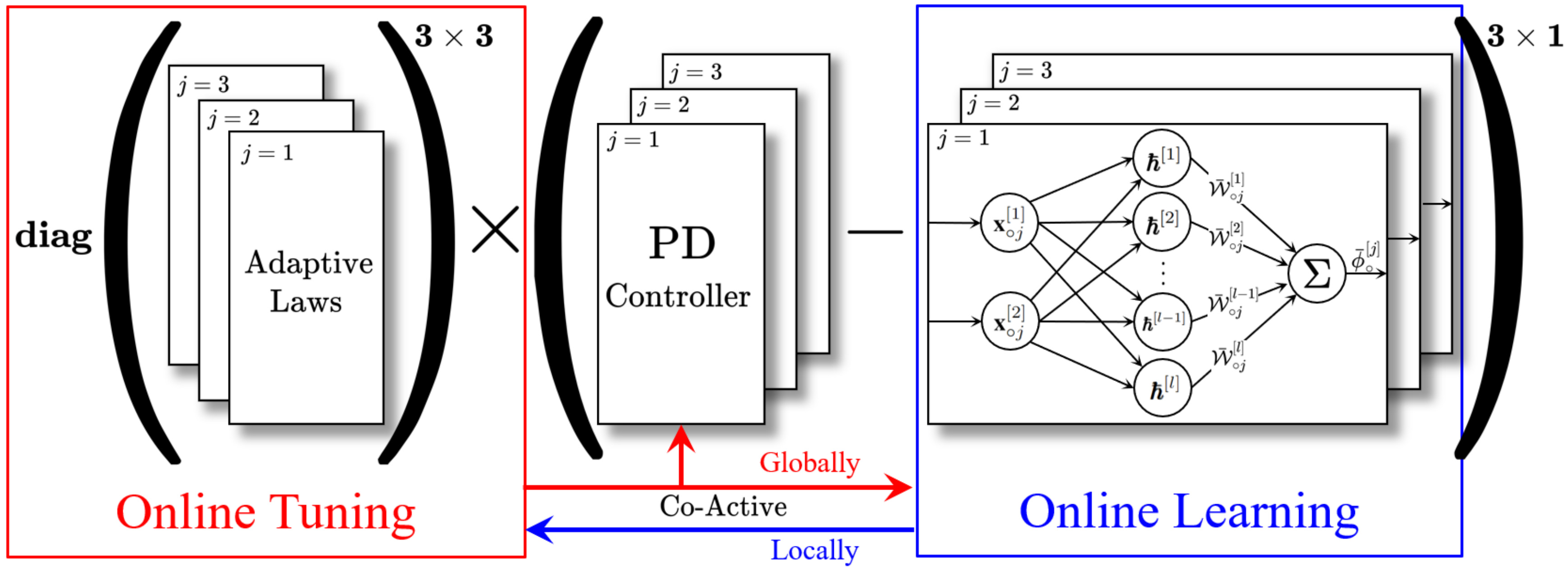}
      \caption{\footnotesize The  structure of  adaptive-neuro control  strategy,  $\bm{\mathcal{U}_{\bm{\circ}}}\in\mathbb{R}^3$ (subscript $\bullet_{\circ\in\{\bm{\textit{x}}, \bm{\textit{R}}\}}$) in Eqs.~(\ref{Translational Adaptive-Neuro Control}) and (\ref{Rotational Adaptive-Neuro Control}). The adaptive model parameters are tuned online and globally scale both the PD controller and the online-learning neural networks. The neural networks, in turn, locally modulate the online tuning process. These  adaptive mechanisms interact with each other. }
      \label{Gao2}
   \end{figure}  
\begin{equation}
{
\footnotesize
\bm{\dot{\bar{m}}}_0^{[j]}:=
\begin{cases}
   \frac{-\bm{\bar{m}}_0^{[j]^2}}{\eta_{\textit{m}}}\bm{\mathcal{E}}_{\bm{\textit{x}} j}^{\top}\bm{P}_j\bm{B}\bm{\mathcal{U}_{\bm{\textit{x}}}}^{[j]}, \,\, \bm{\mathcal{E}}_{\bm{\textit{x}} j}^{\top}\bm{P}_j\bm{B}\bm{\mathcal{U}_{\bm{\textit{x}}}}^{[j]}>0\\[2pt]
 \frac{-\bm{\bar{m}}_0^{[j]^2}}{\eta_{\textit{m}}}\bm{\mathcal{E}}_{\bm{\textit{x}} j}^{\top}\bm{P}_j\bm{B}\bm{\mathcal{U}_{\bm{\textit{x}}}}^{[j]}, \,\, \bm{\mathcal{E}}_{\bm{\textit{x}} j}^{\top}\bm{P}_j\bm{B}\bm{\mathcal{U}_{\bm{\textit{x}}}}^{[j]}\leq 0, \,\,\bm{\bar{m}}_0^{[j]}<\overset{\tiny \text{max}}{m_0}\\[2pt]
 \mathfrak{s}_{\textit{m}}\frac{-\bm{\bar{m}}_0^{[j]^2}}{\eta_{\textit{m}}}, \,\, \,\,\,\,\,\,\,\,\,\,\,\,\,\,\,\,\,\,\,\,\,\,\,\,\,\,\bm{\mathcal{E}}_{\bm{\textit{x}} j}^{\top}\bm{P}_j\bm{B}\bm{\mathcal{U}_{\bm{\textit{x}}}}^{[j]}\leq 0 , \,\,\bm{\bar{m}}_0^{[j]}\geq\overset{\tiny \text{max}}{m_0}
\end{cases}
}
\label{Adaptive Law of mass}
\end{equation}
\begin{equation}
{
\footnotesize
\bm{\dot{\bar{\mathit{J}}}}_0^{[j]}:=
\begin{cases}
   \frac{-\bm{\bar{J}}_0^{[j]^2}}{\eta_{J}}\bm{e}^{[j]}_{\bm{\Omega}_0}\bm{\mathcal{U}_{\bm{\textit{R}}}}^{[j]}, \,\, \,\,\,\,\,\,\,\,\,\,\,\,\,\,\,\,\bm{e}^{[j]}_{\bm{\Omega}_0}\bm{\mathcal{U}_{\bm{\textit{R}}}}^{[j]}>0\\[2pt]
 \frac{-\bm{\bar{J}}_0^{[j]^2}}{\eta_{J}}\bm{e}^{[j]}_{\bm{\Omega}_0}\bm{\mathcal{U}_{\bm{\textit{R}}}}^{[j]}, \,\, \,\,\,\,\,\,\,\,\,\,\,\,\,\,\,\,\bm{e}^{[j]}_{\bm{\Omega}_0}\bm{\mathcal{U}_{\bm{\textit{R}}}}^{[j]}\leq 0, \,\,\bm{\bar{J}}_0^{[j]}<\overset{\tiny \text{max}}{\bm{J}_0}\overset{[j]}{\rule{0pt}{1.5ex}}\\[2pt]
 \mathfrak{s}_{J}\frac{-\bm{\bar{J}}_0^{[j]^2}}{\eta_{J}}, \,\, \,\,\,\,\,\,\,\,\,\,\,\,\,\,\,\,\,\,\,\,\,\,\,\,\,\,\,\,\,\,\,\bm{e}^{[j]}_{\bm{\Omega}_0}\bm{\mathcal{U}_{\bm{\textit{R}}}}^{[j]}\leq 0 , \,\,\bm{\bar{J}}_0^{[j]}\geq\overset{\tiny \text{max}}{\bm{J}_0}\overset{[j]}{\rule{0pt}{1.5ex}}
\end{cases}
}
\label{Adaptive Law of Inertia Tensor}
\end{equation}
with positive constants $\eta_{\textit{m}}$ and $\eta_{J}\in\mathbb{R}$,  preset maximum mass $\overset{\tiny \text{max}}{m_0}\in\mathbb{R}$ and  maximum inertia tensor $\overset{\tiny \text{max}}{\bm{J}_0}\in\mathbb{R}^{3\times3}$ of payload, scaling factors $\mathfrak{s}_{\textit{m}}$ and $\mathfrak{s}_{J}\in\mathbb{R}$, translational error vector along $\bm{\vec{e}}_j$-axis $\bm{\mathcal{E}}_{\bm{\textit{x}} j}:=\left(\bm{e}_{\bm{x}_0}^{[j]}, \bm{\dot{e}}_{\bm{x}_0}^{[j]}\right)^{\top}$, the corresponding Lyapunov matrix $\bm{P}_j\in\mathbb{R}^{2\times2}$ and vector $\bm{B}=\left(0,1\right)^{\top}$.  Additionally, in Eq.~\eqref{Translational Adaptive-Neuro Control},  $\bm{\delta}_{j3}$ is a Kronecker delta
and $\bm{\mathcal{K}}_{\bm{\textit{x}} j}:=\left(\bm{k_{\text{p}}}^{[j]},\bm{k_{\text{d}}}^{[j]}\right)^{\top}$ is the gain vector for translational PD control along $\bm{\vec{e}}_j$-axis  with positive constants $\bm{k_{\text{p}}}$, $\bm{k_{\text{d}}}\in\mathbb{R}^{3}$. In Eq.~\eqref{Rotational Adaptive-Neuro Control},  $k_{R_0}$ and $k_{\Omega_0}\in \mathbb{R}^{+}$ are positive gains  for rotational PD control. $\bm{\bar{\phi}_\bm{\textit{x}}}^{[j]}$ and $\bm{\bar{\phi}_\bm{\textit{R}}}^{[j]}$ are the estimated translational and rotational disturbance dynamics given by Eq.~\eqref{Phi_hat} with estimated weights $\bm{\bar{\mathcal{W}}}_{\bm{\textit{x}} j}$ and $\bm{\bar{\mathcal{W}}}_{\bm{\textit{R}} j}\in\mathbb{R}^l$, adjusted by the following adaptive laws, respectively:
\begin{equation}
    {\footnotesize
    \begin{aligned}
        \bm{\dot{\bar{\mathcal{W}}}}_{\bm{\textit{x}} 
j}:=&\gamma_{\bm{\textit{x}}j}\bm{\mathcal{E}}_{\bm{\textit{x}} j}^{\top}\bm{P}_j\bm{B}\bm{\hbar}(\textbf{x}_{\bm{\textit{x}} j}),
\label{Estimated Weights_x}
    \end{aligned}
    }
\end{equation}
\begin{equation}
    {\footnotesize
    \begin{aligned}
       \bm{\dot{\bar{\mathcal{W}}}}_{\bm{\textit{R}} 
 j}:=&\gamma_{\bm{\textit{R}}j} \bm{e}^{[j]}_{\bm{\Omega}_0}\bm{\hbar}(\textbf{x}_{\bm{\textit{R}} j}),
\label{Estimated Weights_R}
    \end{aligned}
    }
\end{equation}
with corresponding  positive constants $\gamma_{\bm{\textit{x}}j}$ and $\gamma_{\bm{\textit{R}}j}\in\mathbb{R}$. 

Then, the integral compensations are appended to give the  $j^{th}$ element of the enhanced first-level control signals:
\begin{equation}
{\footnotesize
\begin{aligned}
\!\!\!\!\!\!\!\!\!\!\!\!\!\!\!\!\!\!\!\!\!\!\!\!\!\!\!\bm{\mathrm{F}_d}^{[j]}:=\,\, &\bm{\mathcal{U}_{\bm{\textit{x}}}}^{[j]}-\bar{\Delta}_{\bm{x}_0}^{[j]}-\bm{\sum}^n_{i=1}\bar{\Delta}_{\bm{x}_i}^{\parallel[j]},\\[-5pt]
\end{aligned}
}
\label{Payload Translational Control}
\end{equation}
\begin{equation}
{\footnotesize
\begin{aligned}
\bm{\mathrm{M}_d}^{[j]}:=\bm{\mathcal{U}_{\bm{\textit{R}}}}^{[j]}-\bar{\Delta}^{[j]}_{\bm{R}_0}-\bm{\sum}_{i=1}^n\left([\bm{\rho}_i]_{\times}\bm{R}_0^{\top}\bar{\Delta}_{\bm{x}_i}^{\parallel}\right)^{[j]},\\[-5pt]
\end{aligned}
}
\label{Payload Rotational Control}
\end{equation}
where  $\bar{\Delta}_{\bm{x}_0}$, $\bar{\Delta}_{\bm{R}_0}$  and $\bar{\Delta}_{\bm{x}_i}^{\parallel}\in\mathbb{R}^3$ are estimated disturbances from payload and $i^{th}$ quadrotor, which are given by the following integral compensations:
\begin{equation}
{\footnotesize
\begin{aligned}
\dot{\bar{\Delta}}^{[j]}_{\bm{x}_0}:=\frac{\textit{h}_{x_0}}{m'_0}\bm{\mathcal{E}}_{\bm{\textit{x}} j}^{\top}\bm{P}_j\bm{B}, \,\,\,\,\, \dot{\bar{\Delta}}^{[j]}_{\bm{R}_0}:=\frac{\textit{h}_{R_0}}{\bm{J}_0'^{[j]}}\bm{e}_{\bm{\Omega}_0}^{[j]},
\end{aligned}
     }
     \label{integral compensations0}
\end{equation}
\begin{equation}
{\footnotesize
\begin{aligned}
    \dot{\bar{\Delta}}_{\bm{x}_i}:=&\textit{h}_{x_i}(\bm{q}_i \otimes \bm{q}_i)\bigg{\{}\bm{\sum}_{j=1}^3\frac{1}{m'_0}\mathfrak{u}_j\bm{\mathcal{E}}_{\bm{\textit{x}} j}^{\top}\bm{P}_j\bm{B}\\[-5pt]
    &-\bm{J}'^{-1}_0\bm{R}_0[\bm{\rho}_i]_{\times}\bm{e}_{\bm{\Omega}_0}+\frac{\textit{h}_{x_i}}{m_i l_i}[\bm{q}_i]_{\times}(\bm{e}_{\bm{\omega}_i}+\textit{c}_{q}\bm{e}_{\bm{q}_i})\bigg{\}},
\end{aligned}
     }\label{integral compensations}
\end{equation}
in which $m'_0\in\mathbb{R}$ and  $\bm{J}_0'\in\mathbb{R}^3$ are the reference mass and  inertia tensor of payload.   $\textit{c}_{q}$, $\textit{h}_{x_0}$, $\textit{h}_{R_0}$,  $\textit{h}_{x_i}\in\mathbb{R}^{+}$ are positive constants. $\mathfrak{u}_j\in\mathbb{R}^3$ denotes a unit vector with 1 at the $j^{th}$ element and $\bar{\Delta}_{\bm{x}_i}^{\parallel}$ is derived from $\bar{\Delta}_{\bm{x}_i}^{\parallel}=(\bm{q}_i \otimes \bm{q}_i)\bar{\Delta}_{\bm{x}_i}$. 

For the solution of $\bm{P}_j$ and the design of Eqs.~\eqref{Translational Adaptive-Neuro Control}-\eqref{integral compensations} grounded in Lyapunov stability analysis, see Section \ref{Stability Analysis}.
\section{Stability Analysis}
\label{Stability Analysis}
\subsection{Error Dynamics}
The optimal weights of Eq.~\eqref{Phi_hat} and the optimal approximation error between Eq.~\eqref{Phi} and Eq.~\eqref{Phi_hat} are given and defined as follows:
\begin{equation}
    {\footnotesize
    \begin{aligned}
        \bm{\mathcal{W}}_{\circ j}^*\triangleq&\mathrm{arg}\, \underset{\bm{\mathcal{W}}_{\circ j}\in\mathbb{W}_{\circ j}}{\mathrm{min}}\left(\mathrm{sup}\big{\vert}\bm{\phi}_{\circ }^{[j]}-\bm{\bar{\phi}}_{\circ }^{[j]}\big{\vert}\right), \label{W*}
    \end{aligned}
    }
\end{equation}
\begin{equation}
    {\footnotesize
    \begin{aligned}
       \bm{\varpi}^{[j]}_{\circ}\triangleq&\bm{\phi}_{\circ}^{[j]}-\bm{\bar{\phi}}_{\circ}^{[j]}(\textbf{x}_{\circ j}\vert\bm{\mathcal{W}}^*_{\circ j}).
    \label{approximation error}
    \end{aligned}
    }
\end{equation}
\subsubsection{Translational Error Dynamics}
 From Eqs.~\eqref{disturbance-augmented dinamics}, \eqref{Translational Adaptive-Neuro Control}, \eqref{Payload Translational Control}, the translational error dynamics along $\bm{\vec{e}}_j$-axis  is given by:
 \begin{equation}
 {\footnotesize
\begin{aligned}
      \bm{\dot{\mathcal{E}}}_{\bm{\textit{x}} j}={\renewcommand{\arraystretch}{1}\begin{bmatrix}\bm{\dot{e}}_{\bm{x}_0}^{[j]}\\ \bm{\ddot{e}}_{\bm{x}_0}^{[j]}\\\end{bmatrix}}=&\bm{\Lambda}_{\bm{\textit{x}} j}\bm{\mathcal{E}}_{\bm{\textit{x}} j}+\bm{B}\bigg{\{} \widetilde{m}_{j}\bm{\mathcal{U}_{\bm{\textit{x}}}}^{[j]}+\left(\bm{\phi}_{\bm{\textit{x}}}^{[j]}-\bm{\bar{\phi}}_{\bm{\textit{x}}}^{[j]}\right)
      \\&+\frac{1}{m_0}\left(\widetilde{\Delta}_{\bm{x}_0}^{[j]}+\bm{\sum}_{i=1}^n\widetilde{\Delta}_{\bm{x}_i}^{\parallel[j]}\right)+\bm{Y_{\bm{\textit{x}}}}^{[j]}\bigg{\}},
\end{aligned}
}
\label{translational error dynamics1}
\end{equation}
 where  $\widetilde{m}_j\in\mathbb{R}\triangleq\frac{1}{m_0}-\frac{1}{\bm{\bar{m}}_0^{[j]}}$, $\widetilde{\Delta}_{\bm{x}_0}\in\mathbb{R}^3\triangleq\Delta_{\bm{x}_0}-\bar{\Delta}_{\bm{x}_0}$ and $\widetilde{\Delta}_{\bm{x}_i}^{\parallel}\in\mathbb{R}^3\triangleq\Delta_{\bm{x}_i}^{\parallel}-\bar{\Delta}_{\bm{x}_i}^{\parallel}$ denote the  estimation errors,
\begin{equation}
{\footnotesize
\begin{aligned}
     \bm{\Lambda}_{\bm{\textit{x}} j}={\renewcommand{\arraystretch}{1}\begin{bmatrix}0&1\\-\bm{k_{\text{p}}}^{[j]}&-\bm{k_{\text{d}}}^{[j]}\end{bmatrix}}, \,\,\,\,\, \bm{B}={\renewcommand{\arraystretch}{1}\begin{bmatrix}\,0\,\,\\\,1\,\,\end{bmatrix}}.
\end{aligned}
    }
\end{equation}
From Eqs.~ \eqref{Phi_hat},~\eqref{W*} and~\eqref{approximation error} with subscripts $\bullet_{\circ:=\bm{\textit{x}}}$,  Eq.~\eqref{translational error dynamics1} can be further expressed as:
\begin{equation}
{\footnotesize
\begin{aligned}
     \bm{\dot{\mathcal{E}}}_{\bm{\textit{x}} j}=&\bm{\Lambda}_{\bm{\textit{x}} j}\bm{\mathcal{E}}_{\bm{\textit{x}} j}+\bm{B}\bigg{\{} \widetilde{m}_j\bm{\mathcal{U}_{\bm{\textit{x}}}}^{[j]}+\left(\bm{\mathcal{W}}^*_{\bm{\textit{x}} j}-\bm{\bar{\mathcal{W}}}_{\bm{\textit{x}} j}\right)^{\top}\bm{\hbar}(\textbf{x}_{\bm{\textit{x}} j})+\bm{\varpi}^{[j]}_{\bm{\textit{x}}}\\
     &+\frac{1}{m_0}\left(\widetilde{\Delta}_{\bm{x}_0}^{[j]}+\bm{\sum}_{i=1}^n\widetilde{\Delta}_{\bm{x}_{i}}^{\parallel[j]}\right)+\bm{Y_{\bm{\textit{x}}}}^{[j]}\bigg{\}}.
\end{aligned}
    }\label{translational error dynamics2}
\end{equation}

\subsubsection{Rotational Error Dynamics} As noted in \cite{2010 Geometric tracking control of a quadrotor UAV on SE(3)} and \cite{2011 Geometric tracking control of the attitude dynamics of a rigid body on SO(3)}, the rotational error dynamics  is given by:
\begin{equation}
{\footnotesize
\begin{aligned}
    \bm{\dot{e}}_{\bm{R}_0}=
\frac{1}{2}\left(\mathrm{tr}\big{[}\bm{R}_{0}^{\top}\bm{R}_{0_{\bm{d}}}\big{]}\mathrm{I}^{3\times3}-\bm{R}_{0}^{\top}\bm{R}_{0_{\bm{d}}}\right)\bm{e}_{\bm{\Omega}_0},
\end{aligned}
    }\label{rotational error dynamics0}
\end{equation}
\begin{equation}
{\footnotesize
\begin{aligned}
  \bm{\dot{e}}_{\bm{\Omega}_0}=\bm{\dot{\Omega}}_0+[\bm{\Omega}_0]_{\times}\bm{R}_0^{\top}\bm{R}_{0_{\bm{d}}}\bm{\Omega}_{0_{\bm{d}}}-\bm{R}_0^{\top}\bm{R}_{0_{\bm{d}}}\bm{\dot{\Omega}}_{0_{\bm{d}}}.
\end{aligned}
    }\label{rotational error dynamics1}
\end{equation}
From Eqs.~\eqref{disturbance-augmented dinamics}, \eqref{Rotational Adaptive-Neuro Control}, \eqref{Payload Rotational Control} and Eqs.~\eqref{W*}, \eqref{approximation error} with subscripts $\bullet_{\circ:=\bm{\textit{R}}}$, this equation can be further derived in the form of the $j^{th}$ element along $\bm{\vec{b}}_{\bm{0}j}$-axis as:
\begin{equation}
{\footnotesize
\begin{aligned}
 \bm{\dot{e}}^{[j]}_{\bm{\Omega}_0}\!=\!&-k_{R_0}\bm{e}^{[j]}_{\bm{R}_0}\!-\!k_{\Omega_0}\bm{e}^{[j]}_{\bm{\Omega}_0}\!+\!\widetilde{J}_{j}\bm{\mathcal{U}_{\bm{\textit{R}}}}^{[j]}\!+\!\left(\bm{\mathcal{W}}^*_{\bm{\textit{R}} j}\!-\!\bm{\bar{\mathcal{W}}}_{\bm{\textit{R}} j}^{\top}\right)\bm{\hbar}(\textbf{x}_{\bm{\textit{R}} j})\\
&+\bm{\varpi}^{[j]}_{\bm{\textit{R}}}-\left(\bm{J}_0^{-1}[\bm{\Omega}_0]_{\times}\bm{J}_0\bm{\Omega}_0\right)^{[j]}\\
&+\bm{J}_0^{-1[j]}\bigg{\{}\widetilde{\Delta}^{[j]}_{\bm{R}_0}+\bm{\sum}_{i=1}^n\left([\bm{\rho}_i]_{\times}\bm{R}_0^{\top}\widetilde{\Delta}_{\bm{x}_i}^{\parallel}\right)^{[j]}\bigg{\}}+\bm{Y}^{[j]}_{\textit{R}},
\end{aligned}
    }\label{rotational error dynamics2}
\end{equation}
where $\widetilde{J}_{j}\in\mathbb{R}\triangleq\frac{1}{\bm{J}_0^{[j]}}-\frac{1}{\bar{\bm{J}}_0^{[j]}}$, $\widetilde{\Delta}_{\bm{R}_0}\in\mathbb{R}^3\triangleq\Delta_{\bm{R}_0}-\bar{\Delta}_{\bm{R}_0}$ and $\widetilde{\Delta}_{\bm{x}_{i}}^{\parallel}\in\mathbb{R}^3\triangleq\Delta_{\bm{x}_{i}}^{\parallel}-\bar{\Delta}_{\bm{x}_{i}}^{\parallel}$ denote the  estimation errors.
\subsubsection{Cable Attitude Error Dynamics}Since the normal control force $\mathbf{u}_i^{\bot}$ remains unmodified, the cable orientation error dynamics is given as in \cite{2018 Geometric Control of Quadrotor UAVs Transporting
a Cable-Suspended Rigid Body}:
\begin{equation}
\begin{aligned}
{\footnotesize
    -[\bm{q}_i]_{\times}^2\bm{\dot{e}}_{\bm{\omega}_i}=-k_{q}\bm{e}_{\bm{q}_i}-k_{\omega}\bm{e}_{\bm{\omega}_i}-\frac{1}{m_{i}l_{i}}[\bm{q}_i]_{\times}\widetilde{\Delta}_{\bm{x}_{i}}^{\bot},
}
\end{aligned}
\end{equation}
where $k_{q}, k_{\omega}\in\mathbb{R}$ are positive constants, 
 $\bm{e}_{\bm{q}_i}:=\bm{q}_{i_{\bm{d}}} \times \bm{q}_{i}$ and  $\bm{e}_{\bm{\omega}_i}:=\bm{\omega}_{i}+[ \bm{q}_i]^2_{\times}\bm{\omega}_{i_{\bm{d}}}$  are  attitude tracking error vectors with desired angular velocity of $i^{th}$ cable $\bm{\omega}_{i_{\bm{d}}}:=\bm{q}_{i_{\bm{d}}}\times\dot{\bm{q}}_{i_{\bm{d}}}$, and desired direction of $i^{th}$ cable  $\bm{q}_{i_{\bm{d}}}\in \mathbf{S}^2:=-\bm{\mathrm{\mu}}_{i_{\bm{d}}}/\lVert\bm{\mathrm{\mu}}_{i_{\bm{d}}}\lVert$.

\subsection{Stability Proof}
Define the Lyapunov function for complete cable-suspended payload dynamics as $\bm{\mathcal{V}}=\bm{\mathcal{V}}_{\bm{\textit{x}}}+\bm{\mathcal{V}}_{\bm{\textit{R}}}+\bm{\mathcal{V}}_{\bm{q}}+\bm{\mathcal{V}}_{\Delta}$ with the following  Lyapunov candidate terms:
\begin{equation}
{\small
    \begin{aligned}
        \bm{\mathcal{V}}_{\bm{\textit{x}}}=&\bm{\sum}_{j=1}^3\frac{1}{2}\bm{\mathcal{E}}_{\bm{\textit{x}} j}^{\top}\bm{P}_j\bm{\mathcal{E}}_{\bm{\textit{x}} j}+\frac{1}{2}\eta_{\textit{m}}\widetilde{m}_j^2\\
        &+ \frac{1}{2\gamma_{\bm{\textit{x}}j}}\left(\bm{\mathcal{W}}^*_{\bm{\textit{x}} j}-\bm{\bar{\mathcal{W}}}_{\bm{\textit{x}} j}\right)^{\top}\left(\bm{\mathcal{W}}^*_{\bm{\textit{x}} j}-\bm{\bar{\mathcal{W}}}_{\bm{\textit{x}} j}\right),
    \end{aligned}
    }
\end{equation}
\begin{equation}
{\small
    \begin{aligned}
\bm{\mathcal{V}}_{\bm{\textit{R}}}=&k_{R_0}\Psi_{\textit{R}}+\bm{\sum}_{j=1}^3\frac{1}{2}\lVert\bm{e}^{[j]}_{\bm{\Omega}_0}\lVert^2+\frac{1}{2}\eta_{J}\widetilde{J}_j^2
      \\
      &+ \frac{1}{2\gamma_{\bm{\textit{R}}j}}\left(\bm{\mathcal{W}}^*_{\bm{\textit{R}} j}-\bm{\bar{\mathcal{W}}}_{\bm{\textit{R}} j}\right)^{\top}\left(\bm{\mathcal{W}}^*_{\bm{\textit{R}} j}-\bm{\bar{\mathcal{W}}}_{\bm{\textit{R}} j}\right),
\end{aligned}
    }
\end{equation}

\begin{equation}
{\small
    \begin{aligned}
\bm{\mathcal{V}}_{\bm{q}}=\bm{\sum}_{i=1}^{n}\frac{1}{2}\lVert\bm{e}_{\bm{\omega}_i}\lVert^2+k_{q}\Psi_{q_i}+\textit{c}_{q}\bm{e}_{\bm{q}_i}\cdot\bm{e}_{\bm{\omega}_i},
\end{aligned}
    }
\end{equation}
\begin{equation}
{\small
    \begin{aligned}
    \bm{\mathcal{V}}_{\Delta}=\frac{1}{2\textit{h}_{x_0}}\lVert\widetilde{\Delta}_{\bm{x}_0}\lVert^2+\frac{1}{2\textit{h}_{R_0}}\lVert\widetilde{\Delta}_{\bm{R}_0}\lVert^2+\bm{\sum}_{i=1}^{n}\frac{1}{2\textit{h}_{x_i}}\lVert\widetilde{\Delta}_{\bm{x}_i}\lVert^2,
\end{aligned}
}
\end{equation}
where $\eta_{\textit{m}}$, $\eta_{J}$, $\gamma_{\bm{\textit{x}}j}$ and $\gamma_{\bm{\textit{R}}j}\in\mathbb{R}$  are positive constants. $\Psi_{\textit{R}}\in\mathbb{R}^{+}\triangleq\frac{1}{2}\mathrm{tr}[\mathrm{I}^{3\times3}-\bm{R}_{0_{\bm{d}}}^{\top}\bm{R}_0]$, $\Psi_{q_i}\in\mathbb{R}^{+}\triangleq1-\bm{q}_{i}\cdot\bm{q}_{i_{\bm{d}}}$ and $\bm{P}_j\in\mathbb{R}^{2\times2}$ is a symmetric positive-definite matrix that follows the Lyapunov equation with matrix $\bm{Q}_j>0$:  $\bm{\Lambda}_{\textit{x}j}^{\top}\bm{P}_j+\bm{P}_j\bm{\Lambda}_{\textit{x}j}=-\bm{Q}_j$. As noted in \cite{2018 Geometric Control of Quadrotor UAVs Transporting
a Cable-Suspended Rigid Body}, if the positive constant $\textit{c}_{q}$ is sufficiently small, $\bm{\mathcal{V}}_{\bm{q}}$ is  positive-definite. Since $\bm{\mathcal{V}}_{\textit{x}}$, $\bm{\mathcal{V}}_{\textit{R}}$ and $\bm{\mathcal{V}}_{\Delta}\geq0$,  it follows that $\bm{\mathcal{V}}$ is also positive-definite.

The time-derivative of the complete Lyapunov function can be then given by $\bm{\dot{\mathcal{V}}}=\bm{\dot{\mathcal{V}}}_{\bm{\textit{x}}}+\bm{\dot{\mathcal{V}}}_{\bm{\textit{R}}}+\bm{\dot{\mathcal{V}}}_{\bm{q}}+\bm{\dot{\mathcal{V}}}_{\Delta}$ with:
\begin{equation}
    {\footnotesize
    \begin{aligned}
\bm{\dot{\mathcal{V}}}_{\bm{\textit{x}}}=&\bm{\sum}_{j=1}^3-\frac{1}{2}\bm{\mathcal{E}}_{\bm{\textit{x}} j}^{\top}\bm{Q}_j\bm{\mathcal{E}}_{\bm{\textit{x}} j}
      +\widetilde{m}_j\left(\bm{\mathcal{E}}_{\bm{\textit{x}} j}^{\top}\bm{P}_j\bm{B} \bm{\mathcal{U}}_{\textit{x}}^{[j]}+\eta_{\textit{m}}\frac{\bm{\dot{\bar{m}}}_0^{[j]}}{\bm{\bar{m}}_0^{[j]^2}}\right)\\
       &+\frac{1}{\gamma_{\textit{x}j}}\left(\bm{\mathcal{W}}^*_{\bm{\textit{x}} j}-\bm{\bar{\mathcal{W}}}_{\bm{\textit{x}} j}\right)^{\top}\left(\gamma_{\textit{x}j}\bm{\mathcal{E}}_{\bm{\textit{x}} j}^{\top}\bm{P}_j\bm{B}\bm{\hbar}(\textbf{x}_{\textit{x}j})-\bm{\dot{\bar{\mathcal{W}}}}_{\textit{x}j}\right)\\
       &+\bm{\mathcal{E}}_{\bm{\textit{x}} j}^{\top}\bm{P}_j\bm{B}\Bigg{\{}\bm{\varpi}^{[j]}_{\textit{x}}+\frac{1}{m_0}\left(\widetilde{\Delta}_{\bm{x}_0}^{[j]}+\bm{\sum}_{i=1}^n\widetilde{\Delta}_{\bm{x}_i}^{\parallel[j]}\right)+\bm{Y}_{\textit{x}}^{[j]}\Bigg{\}}, 
    \end{aligned}
    }
\end{equation}
\begin{equation}
    {\footnotesize
    \begin{aligned}
\!\!\!\!\!\!\!\!\!\!\bm{\dot{\mathcal{V}}}_{\bm{\textit{R}}}=&k_{R_0}\dot{\Psi}_{\textit{R}}+\bm{\sum}_{j=1}^3-k_{R_0}\bm{e}^{[j]}_{\bm{\Omega}_0}\bm{e}^{[j]}_{\bm{R}_0}-k_{\Omega_0}\lVert\bm{e}^{[j]}_{\bm{\Omega}_0}\lVert^2\\
       &+\frac{1}{\gamma_{\bm{\textit{R}}j}}\left(\bm{\mathcal{W}}^*_{\bm{\textit{R}} j}-\bm{\bar{\mathcal{W}}}_{\bm{\textit{R}} j}\right)^{\top}\left(\gamma_{\bm{\textit{R}}j}\bm{e}^{[j]}_{\bm{\Omega}_0}\bm{\hbar}(\textbf{x}_{\textit{R}j})-\bm{\dot{\bar{\mathcal{W}}}}_{\textit{R}j}\right)\\[-10pt]
    \end{aligned}\notag
    }
\end{equation}

\begin{equation}
    {\footnotesize
    \begin{aligned}
       &\!\!\!+\widetilde{J}_j
\left(\bm{e}^{[j]}_{\bm{\Omega}_0}\bm{\mathcal{U}}_{\textit{R}}^{[j]}+\eta_{J}\frac{\bm{\dot{\bar{\mathit{J}}}}_0^{[j]}}{\bm{\bar{J}}_0^{[j]^2}}\right)+\bm{e}^{[j]}_{\bm{\Omega}_0}\Bigg{\{}\!\!\!-\!\left(\bm{J_0}^{-1}[\bm{\Omega}_0]_{\times}\bm{J}_0\bm{\Omega}_0\right)^{[j]}\\
&\!\!\!+\bm{\varpi}^{[j]}_{\textit{R}}+\frac{1}{\bm{J}_0^{[j]}}\bigg{\{}\widetilde{\Delta}^{[j]}_{\bm{R}_0}+\bm{\sum}_{i=1}^n\left([\bm{\rho}_i]_{\times}\bm{R}_0^{\top}\widetilde{\Delta}_{\bm{x}_i}^{\parallel}\right)^{[j]}\bigg{\}}+\bm{Y}_{\textit{R}}^{[j]}\Bigg{\}}.
    \end{aligned}
    }
\end{equation}
\begin{equation}
    {\footnotesize
    \begin{aligned}
      \bm{\dot{\mathcal{V}}}_{\bm{q}}=&\bm{\sum}_{i=1}^{n}-(k_{\omega}-\textit{c}_{q})\lVert\bm{e}_{\bm{\omega}_i}\lVert^2-\textit{c}_{q}k_{q}\lVert\bm{e}_{\bm{q}_i}\lVert^2\\&-\textit{c}_{q}k_{\omega}\bm{e}_{\bm{q}_i}\cdot\bm{e}_{\bm{\omega}_i}-\left(\bm{e}_{\bm{\omega}_i}+\textit{c}_{q}\bm{e}_{\bm{q}_i}\right)\cdot\frac{[\bm{q}_i]_{\times}}{m_{i}l_{i}}\widetilde{\Delta}_{\bm{x}_{i}}^{\bot},
    \end{aligned}
    }
\end{equation}
\begin{equation}
    {\footnotesize
    \begin{aligned}
    \bm{\dot{\mathcal{V}}}_{\Delta}=&-\frac{1}{\textit{h}_{x_0}}\widetilde{\Delta}_{\bm{x}_0}\cdot\dot{\bar{\Delta}}_{\bm{x}_0}-\frac{1}{\textit{h}_{R_0}}\widetilde{\Delta}_{\bm{R}_0}\cdot\dot{\bar{\Delta}}_{\bm{R}_0}\\
     &-\bm{\sum}_{i=1}^{n}\frac{1}{\textit{h}_{x_i}}\left(\widetilde{\Delta}^{\parallel}_{\bm{x}_{i}}\cdot\dot{\bar{\Delta}}^{\parallel}_{\bm{x}_{i}}+\widetilde{\Delta}^{\bot}_{\bm{x}_{i}}\cdot\dot{\bar{\Delta}}^{\bot}_{\bm{x}_{i}}\right).
    \end{aligned}
    }
\end{equation}

Design the $\bm{\dot{\bar{m}}}_0^{[j]}$, $\bm{\dot{\bar{\mathit{J}}}}_0^{[j]}$, $\bm{\dot{\bar{\mathcal{W}}}}_{\bm{\textit{x}} 
j}$, $\bm{\dot{\bar{\mathcal{W}}}}_{\bm{\textit{R}} 
j}$, $\dot{\bar{\Delta}}_{\bm{x}_0}$, $\dot{\bar{\Delta}}_{\bm{R}_0}$ and $\dot{\bar{\Delta}}_{\bm{x}_{i}}$ as given in Eqs.~\eqref{Adaptive Law of mass}-\eqref{Estimated Weights_R}, \eqref{integral compensations0} and  \eqref{integral compensations}.   Then, with the fact that $\dot{\Psi}_{\textit{R}}=\bm{e}_{\bm{R}_0}\cdot\bm{e}_{\bm{\Omega}_0}$ \cite{2010 Geometric tracking control of a quadrotor UAV on SE(3)}, the time-derivative of the complete Lyapunov function $\bm{\dot{\mathcal{V}}}$ reduces to :
\begin{equation}
{\footnotesize
    \begin{aligned}
\bm{\dot{\mathcal{V}}}=&\Bigg{(}\bm{\sum}_{j=1}^3-\frac{1}{2}\bm{\mathcal{E}}_{\bm{\textit{x}} j}^{\top}\bm{Q}_j\bm{\mathcal{E}}_{\bm{\textit{x}} j}+\bm{\mathcal{E}}_{\bm{\textit{x}}j}^{\top}\bm{P}_j\bm{B}\bm{\Xi}^{[j]}_{\bm{\textit{x}}}-k_{\Omega_0}\lVert\bm{e}^{[j]}_{\bm{\Omega}_0}\lVert^2+\\
&\bm{e}^{[j]}_{\bm{\Omega}_0}\bm{\Xi}^{[j]}_{\bm{\textit{R}}}\Bigg{)}-{\bm{e}_{\bm{\Omega}_0}\cdot\left(\bm{J}_0^{-1}[\bm{\Omega}_0]_{\times}\bm{J}_0\bm{\Omega}_0\right)}-\bm{\sum}_{i=1}^n\bm{\mathcal{E}}_{\bm{q}_i}^{\top}\bm{\mathcal{Z}}\bm{\mathcal{E}}_{\bm{q}_i},
\end{aligned}
}
\label{dot V}
\end{equation}
where $\bm{\mathcal{E}}_{\bm{q}_i}=\Big{[}\lVert\bm{e}_{\bm{q}_i}\lVert, \lVert\bm{e}_{\bm{\omega}_i}\lVert\Big{]}^{\top}$, $\bm{\mathcal{Z}}={\begin{bmatrix}
         \textit{c}_{q}k_{q} &\frac{\textit{c}_{q}k_{\omega}}{2}\\
\frac{\textit{c}_{q}k_{\omega}}{2}&k_{\omega}\text{-}\textit{c}_{q}
\end{bmatrix}}$ and
{\footnotesize
\begin{align}
&\begin{aligned}
&\bm{\Xi}^{[j]}_{\bm{\textit{x}}}=\bm{\varpi}^{[j]}_{\textit{x}}+\left(\frac{1}{m_0}-\frac{1}{m'_0}\right)\left(\widetilde{\Delta}_{\bm{x}_0}^{[j]}+\bm{\sum}_{i=1}^n\widetilde{\Delta}_{\bm{x}_i}^{\parallel[j]}\right)+\bm{Y}_{\textit{x}}^{[j]},\\
&\bm{\Xi}^{[j]}_{\bm{\textit{R}}}=\bm{\varpi}^{[j]}_{\textit{R}}\!+\!\left(\frac{1}{\bm{J}_0^{[j]}}-\frac{1}{\bm{J}_0'^{[j]}}\right)\bigg{\{}\widetilde{\Delta}^{[j]}_{\bm{R}_0}\!\!+\!\!\bm{\sum}_{i=1}^n\left([\bm{\rho}_i]_{\times}\bm{R}_0^{\top}\widetilde{\Delta}_{\bm{x}_i}^{\parallel}\right)^{[j]}\!\!\!\bigg{\}}\!+\!\bm{Y}_{\textit{R}}^{[j]}.
\end{aligned}\notag
\end{align}
}

In the case $\bm{\Omega}_{0_{\bm{d}}}=0$, the fifth term of Eq.~\eqref{dot V} vanishes because  $\bm{e}_{\bm{\Omega}_0}=\bm{\Omega}_0$. Additionally, if the reference model parameters are matched with plant: $m'_0=m_0$, $\bm{J}'_0=\bm{J}_0$,  the second terms of $\bm{\Xi}^{[j]}_{\bm{\textit{x}}}$ and $\bm{\Xi}^{[j]}_{\bm{\textit{R}}}$ vanish. 
Given that the optimal approximation error $\lVert\bm{\varpi}_{\textit{x}}\lVert$ and $\lVert\bm{\varpi}_{\textit{R}}\lVert$ are bounded, and the $\lVert\bm{Y}_{\textit{x}}\lVert$ and $\lVert\bm{Y}_{\textit{R}}\lVert$ are also bounded \cite{2018 Geometric Control of Quadrotor UAVs Transporting
a Cable-Suspended Rigid Body}, it follows that $\lVert\bm{\Xi}_{\textit{x}}\lVert$ and $\lVert\bm{\Xi}_{\textit{R}}\lVert$ are likewise bounded. Under the foregoing preconditions, it holds that:
\begin{equation}
    {\footnotesize
    \begin{aligned}
     \bm{\dot{\mathcal{V}}}\leq&\bm{\sum}_{j=1}^3-\frac{1}{2}\lVert\bm{\mathcal{E}}_{\bm{\textit{x}} j}\lVert\Big{\{}\underset{\tiny\text{min}}{\lambda}\left(\bm{Q}_j\right)\lVert\bm{\mathcal{E}}_{\bm{\textit{x}} j}\lVert-2\lVert\overset{\tiny\text{max}}{\bm{\Xi}^{[j]}_{\bm{\textit{x}}}}\lVert\overset{\tiny\text{max}}{\lambda}\left(\bm{P}_j\right)  \Big{\}}\\
     &-\lVert\bm{e}^{[j]}_{\bm{\Omega}_0}\lVert\left(k_{\Omega_0}\lVert\bm{e}^{[j]}_{\bm{\Omega}_0}\lVert-\bm{\Xi}^{[j]}_{\bm{\textit{R}}}\right)-\bm{\sum}_{i=1}^n\bm{\mathcal{E}}_{\bm{q}_i}^{\top}\bm{\mathcal{Z}}\bm{\mathcal{E}}_{\bm{q}_i}.
    \end{aligned}
    }\label{dot V<=}
\end{equation}

Design eigenvalue $\underset{\tiny\text{min}}{\lambda}\left(\bm{Q}_j\right)\geq\frac{2\lVert\overset{\tiny\text{max}}{\bm{\Xi}^{[j]}_{\bm{\textit{x}}}}\lVert\overset{\tiny\text{max}}{\lambda}\left(\bm{P}_j\right)}{\lVert\bm{\mathcal{E}}_{\bm{\textit{x}} j}\lVert}$ , $k_{\Omega_0}\geq\frac{\bm{\Xi}^{[j]}_{\bm{\textit{R}}}}{\lVert\bm{e}^{[j]}_{\bm{\Omega}_0}\lVert}$ and choose $\textit{c}_{q}$ to be sufficiently small such that the matrix $\bm{\mathcal{Z}}$ is positive-definite. The proposed control system then achieved semi-global practical stability. 

When $\bm{\Omega}_{0_{\bm{d}}}\ne0$ and the model parameters are unmatched, the  dynamical tracking stability is not guaranteed since the controllers lack the  knowledge of payload mass and inertia tensor. Nevertheless, as long as the $\underset{\tiny\text{min}}{\lambda}\left(\bm{Q}_j\right)$ and $k_{\Omega_0}$ are appropriate and sufficiently large, the system can still achieve stability over a wide range of model uncertainties.
The  enhanced robustness is demonstrated  in Section \ref{numerical simulation}.
\begin{figure}[t]
\vspace{-0em} 
  \centering
  \begin{subfigure}[b]{0.238\textwidth}
    \centering
    \includegraphics[width=\linewidth]{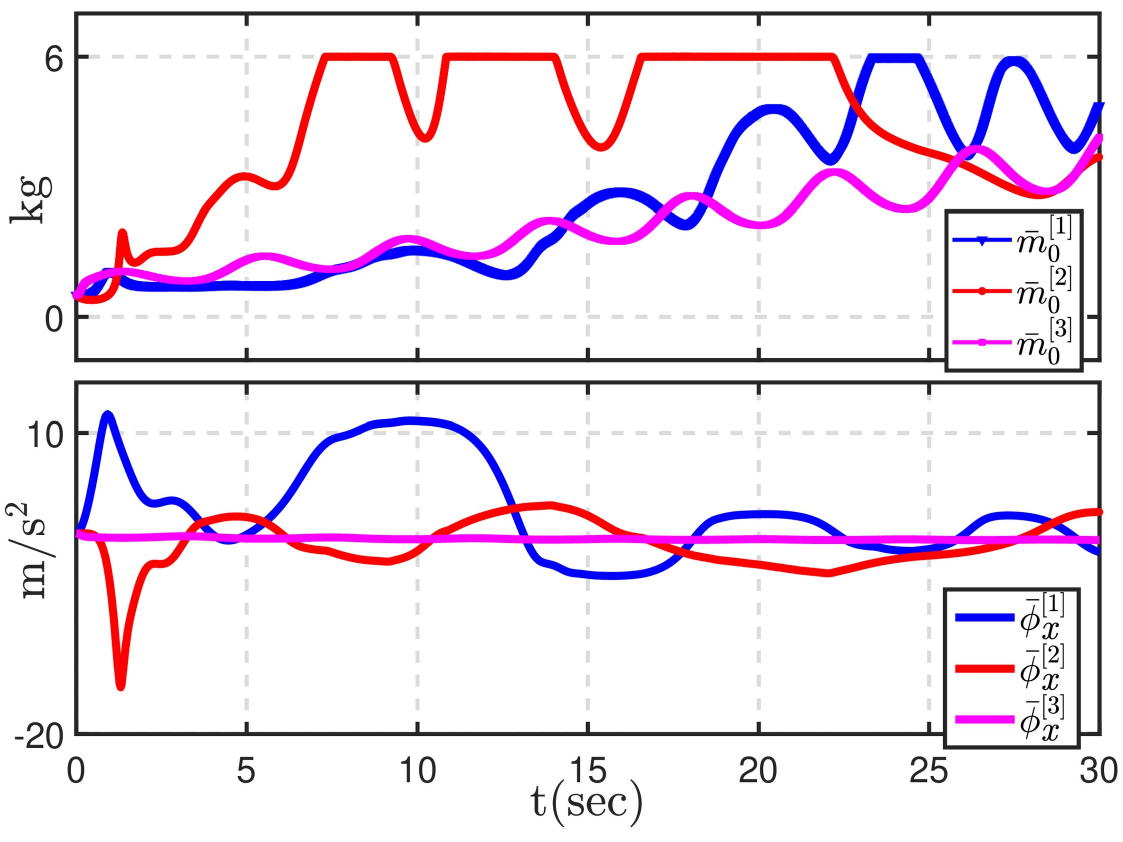}
    \caption{\footnotesize Values of $\bm{\bar{m}}_0^{[j]}$ and $\bm{\bar{\phi}}_{\textit{x}}^{[j]}$ }
    \label{Gao3_a}
  \end{subfigure}
  \hspace{0mm}
  \begin{subfigure}[b]{0.238\textwidth}
    \centering
    \includegraphics[width=\linewidth]{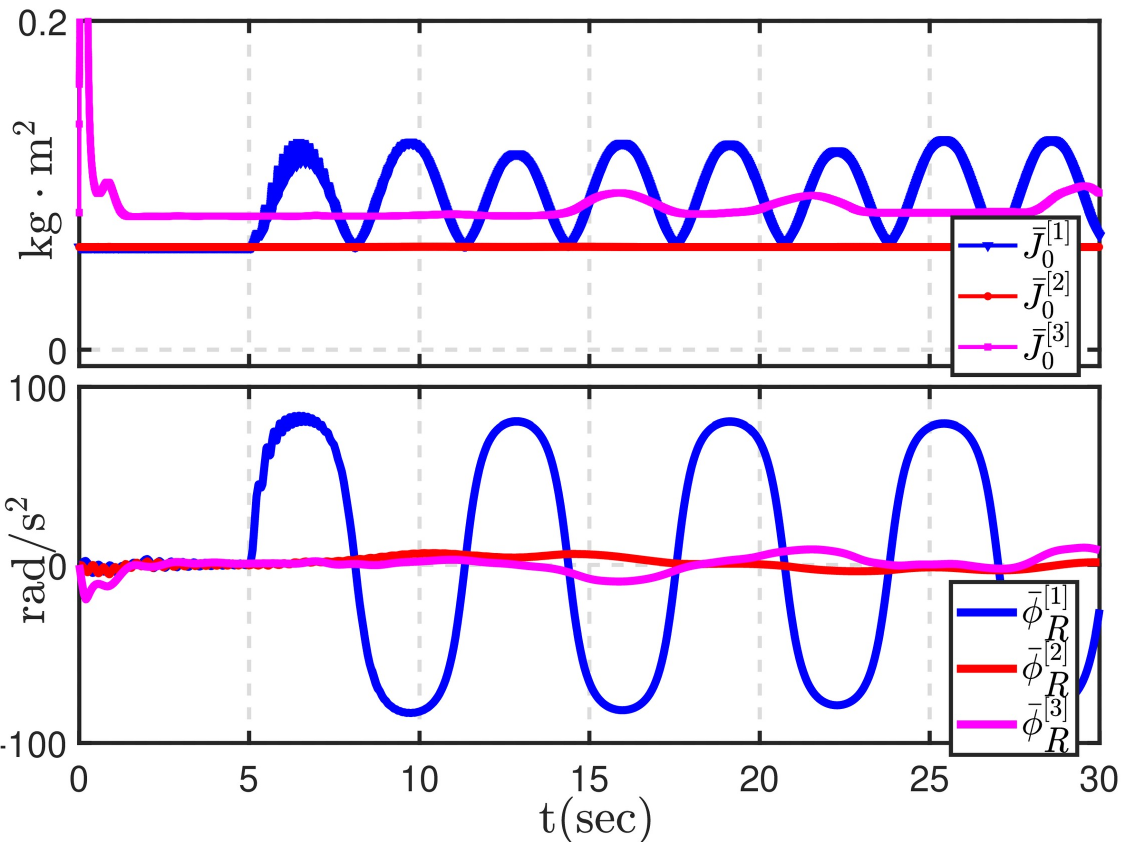}
    \caption{\footnotesize Values of $\bm{\bar{J}}_0^{[j]}$ and $\bm{\bar{\phi}}_{\textit{R}}^{[j]}$  }
    \label{Gao3_b}
  \end{subfigure}
  \vspace{-1em} 
   \caption{\footnotesize The values of estimated model parameters $\bm{\bar{m}}_0^{[j]}$, $\bm{\bar{J}}_0^{[j]}$   tuned by adaptive laws, and disturbance dynamics $\bm{\bar{\phi}}_{\textit{x}}^{[j]}$, $\bm{\bar{\phi}}_{\textit{R}}^{[j]}$ learned by neural networks during simulations in Fig.~\ref{Gao4}. Note that due to the design of adaptive laws in Eqs.~(\ref{Adaptive Law of mass}) and (\ref{Adaptive Law of Inertia Tensor}), the parameters  $\bm{\bar{m}}_0^{[j]}$and $\bm{\bar{J}}_0^{[j]}$ are bounded to the preset maximum values: $\overset{\tiny \text{max}}{m_0}$= 6$\mathrm{kg}$ and $\overset{\tiny \text{max}}{\bm{J}_0}$ = $\mathbf{diag}[0.75,0.75,1](\mathrm{kg}\cdot\mathrm{m}^2)$. }
  \label{Gao3}
  \vspace{-1.5em} 
\end{figure}
 \vspace{-0.5em} 
\section{Numerical Simulation}
\label{numerical simulation}

\begin{figure*}[htbp] 
\vspace{-0em} 
  \centering
  \begin{subfigure}[b]{0.3\textwidth}
    \centering
    \includegraphics[width=\linewidth]{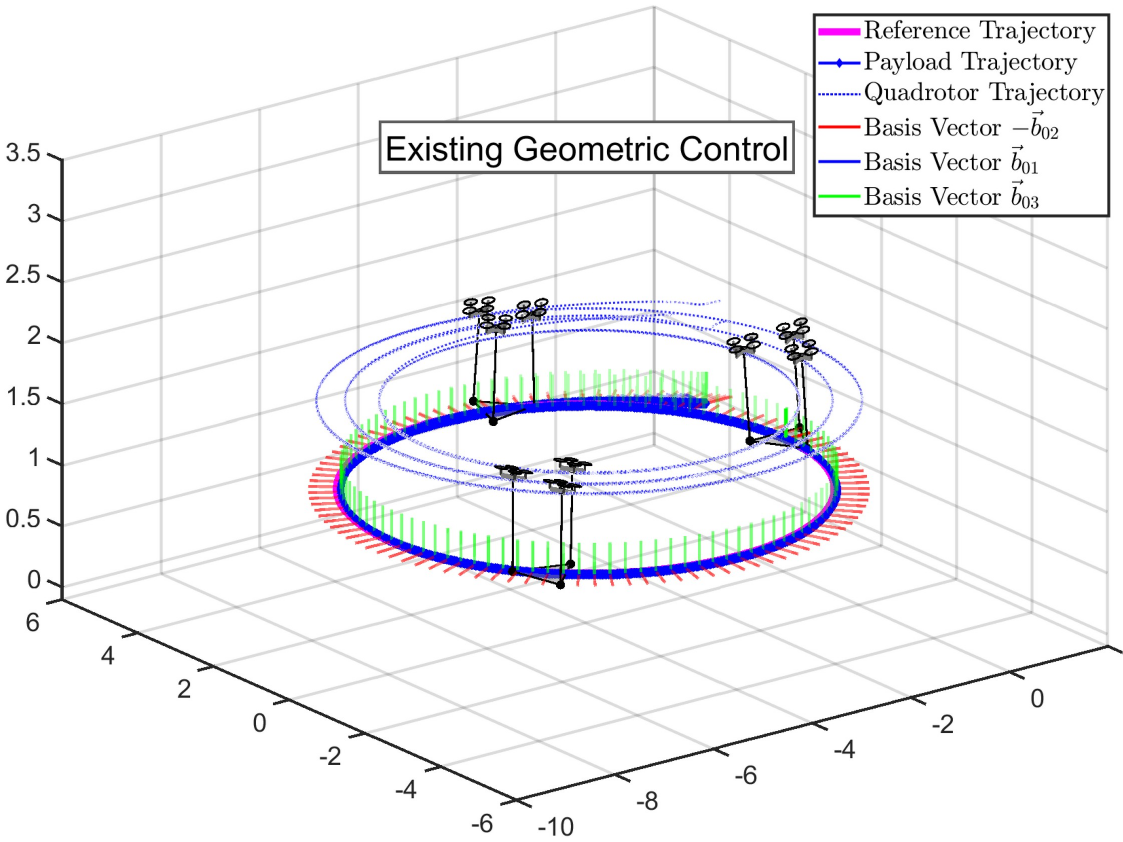}
    \captionsetup{labelformat=empty, labelsep=none} 
    \caption*{}
  \end{subfigure}
  \hspace{2mm}
  \begin{subfigure}[b]{0.3\textwidth}
    \centering
    \includegraphics[width=\linewidth]{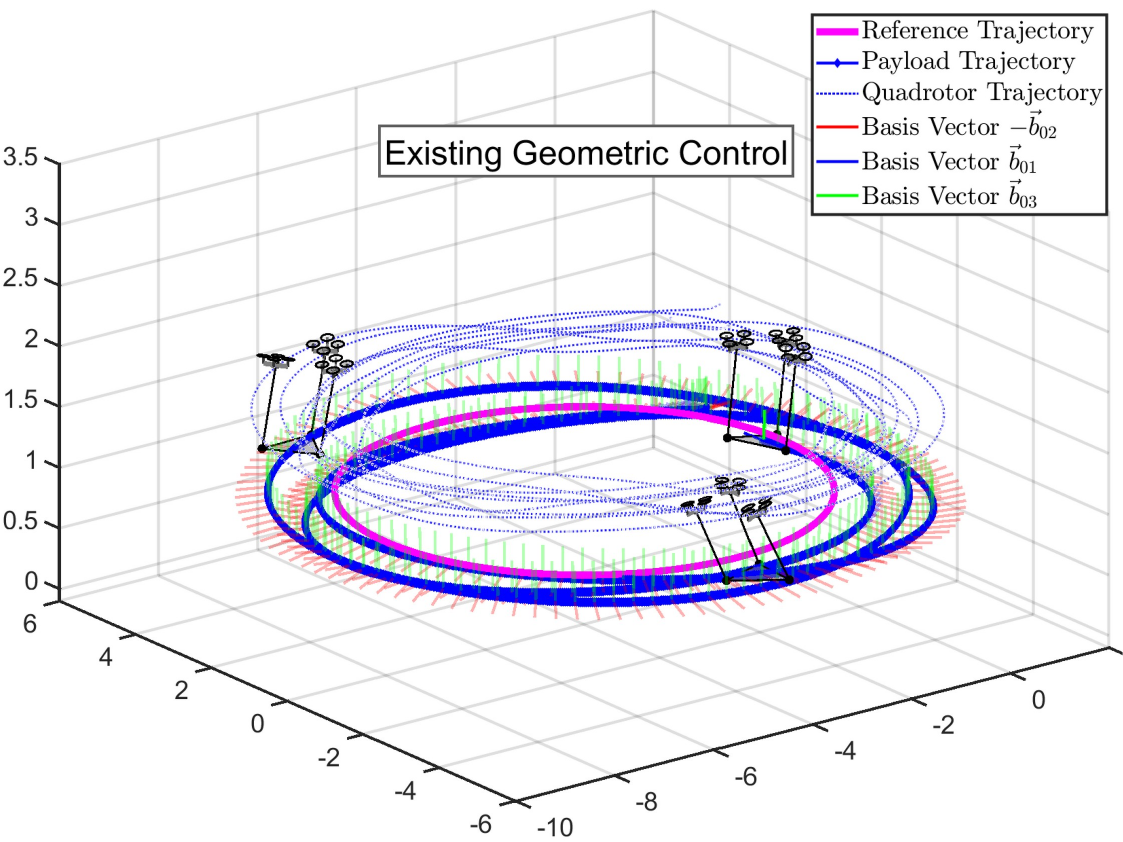}
    \captionsetup{labelformat=empty, labelsep=none} 
    \caption*{}
  \end{subfigure}
  \hspace{2mm}
  \begin{subfigure}[b]{0.3\textwidth}
    \centering
    \includegraphics[width=\linewidth]{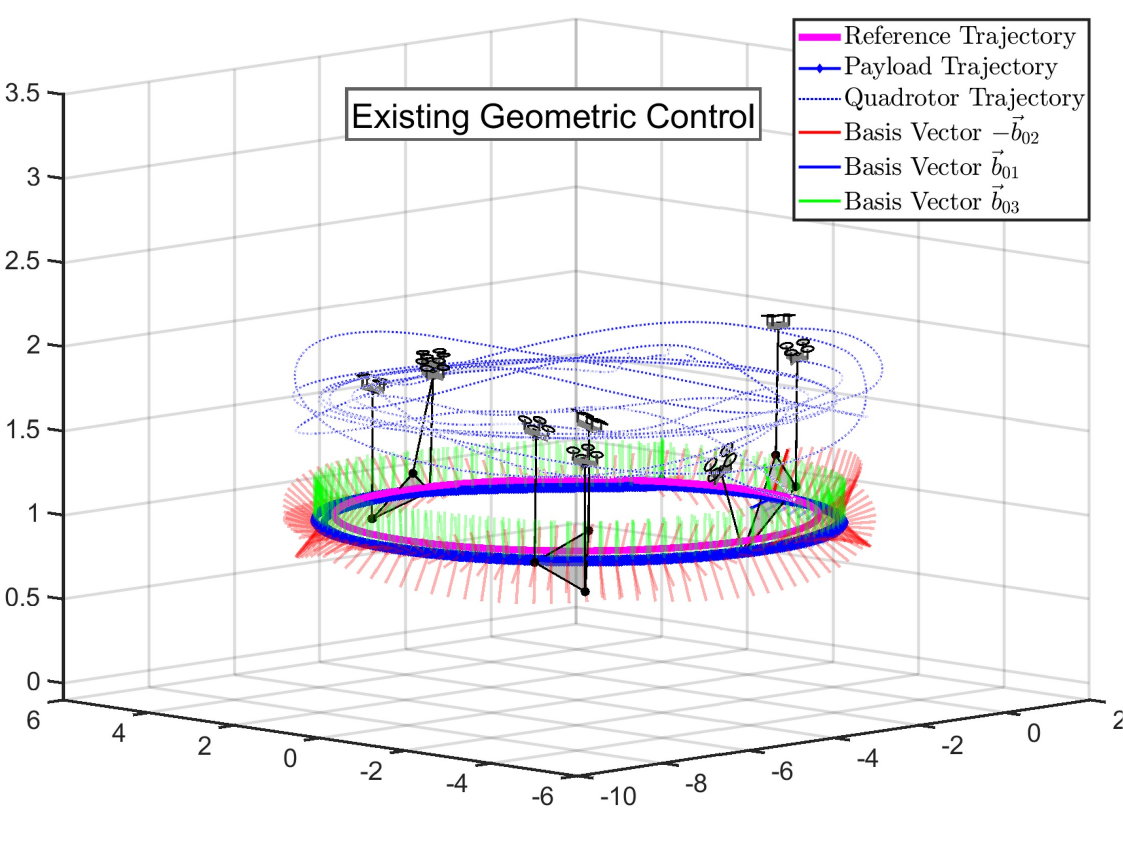}
    \captionsetup{labelformat=empty, labelsep=none} 
    \caption*{}
  \end{subfigure}
    \vspace{-2em} 

  \captionsetup{labelformat=default, labelsep=colon}
  \begin{subfigure}[b]{0.3\textwidth}
    \centering
    \includegraphics[width=\linewidth]{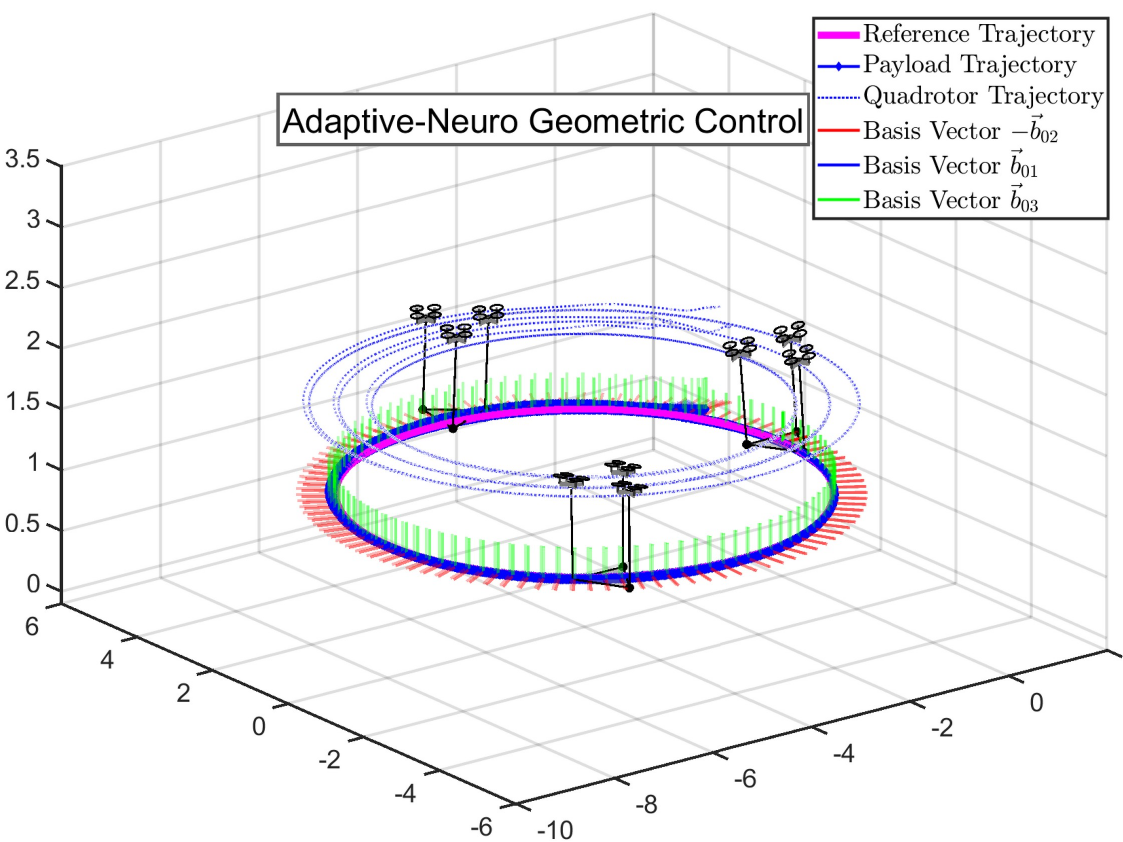}
    \caption{\footnotesize Comparison group A  }
    \label{Gao4_a}
  \end{subfigure}
  \hspace{2mm}
  \begin{subfigure}[b]{0.3\textwidth}
    \centering
    \includegraphics[width=\linewidth]{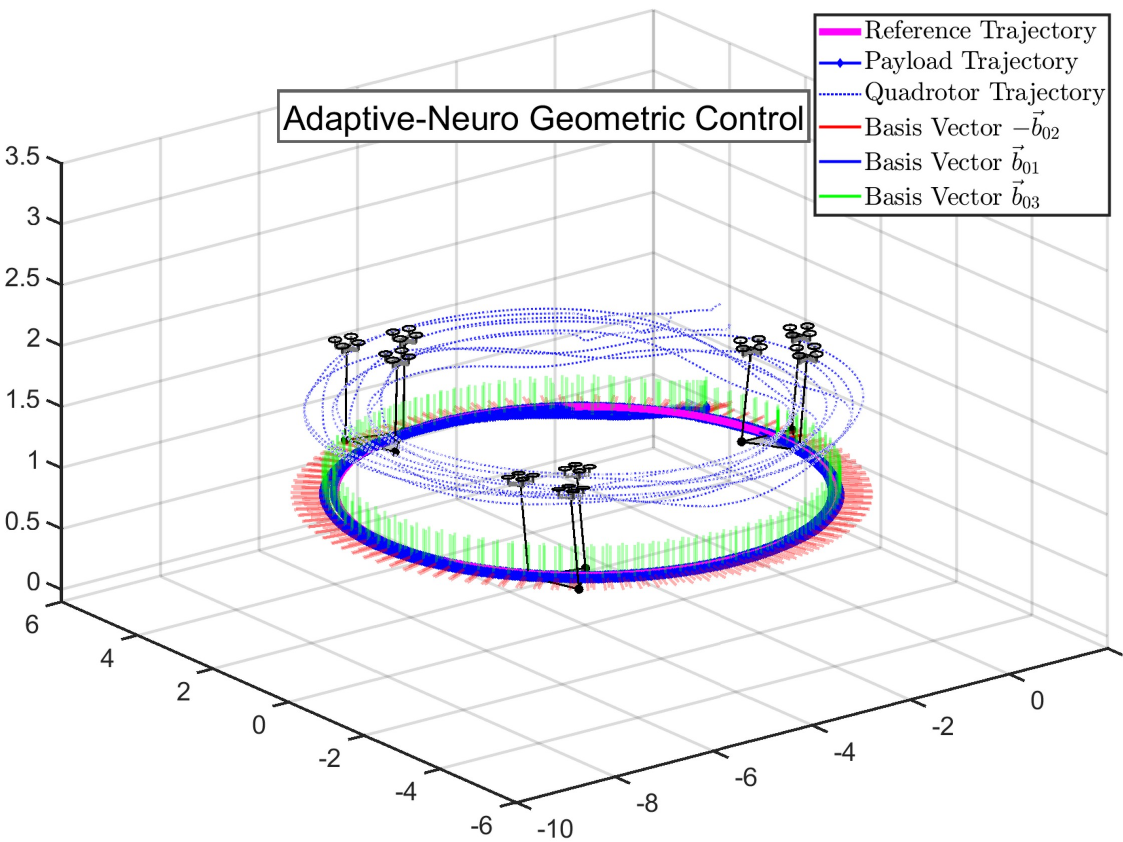}
    \caption{\footnotesize  Comparison group B}
    \label{Gao4_b}
  \end{subfigure}
  \hspace{2mm}
  \begin{subfigure}[b]{0.3\textwidth}
    \centering
    \includegraphics[width=\linewidth]{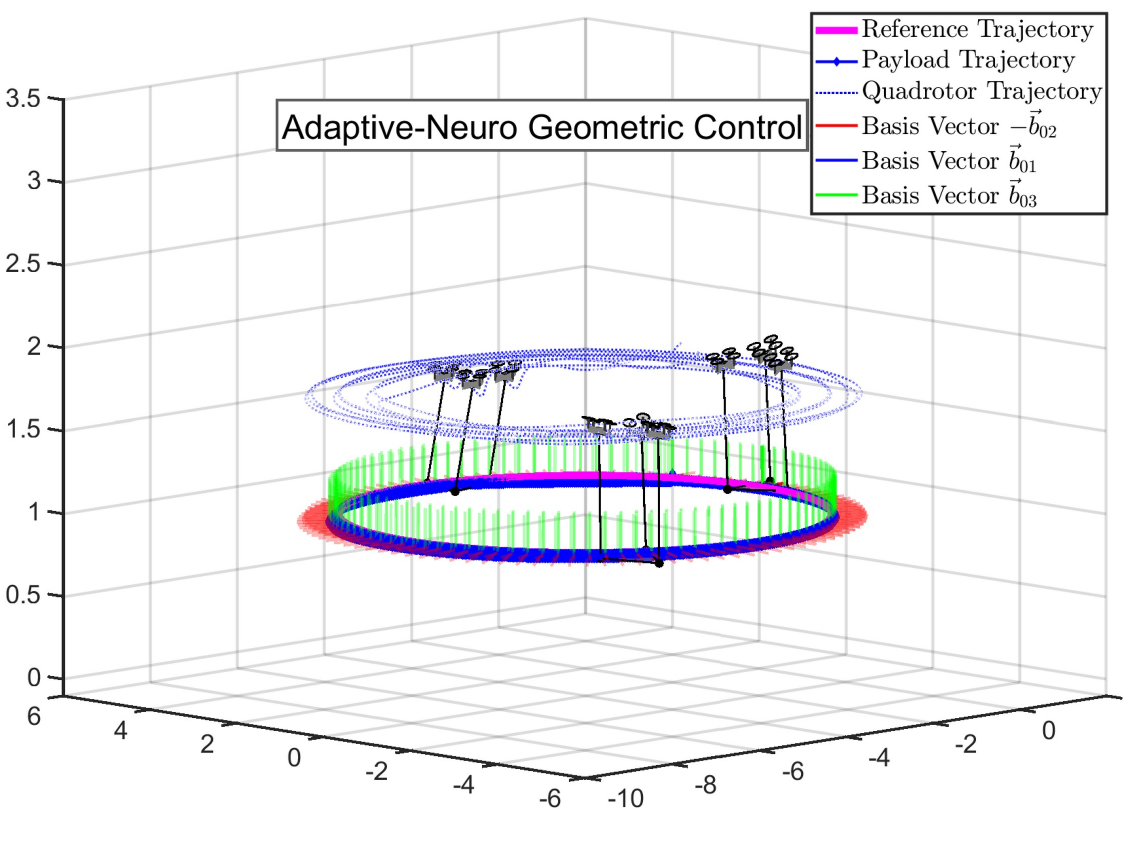}
    \caption{\footnotesize   Comparison group C}
    \label{Gao4_c}
  \end{subfigure}
  \caption{\footnotesize Simulation results of 3-quadrotor transportation in MATLAB Simulink (ODE3 Bogacki–Shampine solver). Group A ensured the appropriateness of the preset controller parameters. In group B, the adaptive-neuro geometric control showed enhanced translational robustness in the presence of parametric uncertainties and irregular disturbance forces   $\Delta^{B}_{\bm{x}_0}$=$[15\sin(\sin(0.02t)t)$+$\cos(0.5t),15\sin(\cos(0.04t$+$\pi)t)$+$5\cos(0.5t),$-$25\sin(1.5t)$+$\cos(0.5t)]^{\top}(\mathrm{N})$. Meanwhile, group C presented enhanced rotational robustness in the presence of parametric uncertainties and a strong disturbance moment $\Delta^{C}_{\bm{R}_0}$= $[10\sin(t$-$5),0,0]^{\top}(\mathrm{Nm})$ along the $b_{01}$ axis from $t $ = 5s. For simulation video, refer to \url{https://staff.aist.go.jp/kamimura.a/ACC/video.mp4}.}
   \vspace{-1em} 
  \label{Gao4}
\end{figure*}
 \vspace{-0em} 

Three comparison groups A, B and C, as shown in Fig.~\ref{Gao4_a}, \ref{Gao4_b}, \ref{Gao4_c} respectively, were conducted to demonstrate the enhanced robustness of our algorithm  over the existing method. During the whole simulations, we ensured that the adaptive-neuro geometric control (our algorithm) and the existing geometric control shared the same setup, such as the preset reference parameters: $\bm{J}_0'$=$\mathbf{diag}[\frac{1}{8},\frac{1}{8},\frac{1}{6}](\mathrm{kg}\cdot\mathrm{m}^2)$, $m'_0$=$1(\mathrm{kg})$; PD controller gains:  $\bm{k_{\text{p}}}$=$[20,20,1000]^{\top}$, $\bm{k_{\text{d}}}$=$[10,10,200]^{\top}$, $k_{R_0}$=20, $k_{\Omega_0}$=10; integral compensation gains:   $\textit{c}_{q}$=0.01, $\textit{h}_{x_0}$=1,  $\textit{h}_{R_0}$= $\textit{h}_{x_i}$=0.1. The parameters of adaptive-neuro geometric control are selected as: hidden layer neurons $l$=$5$, width $b_k\in[1,2]$,  $\eta_{\textit{m}}$=$\eta_{J}$ =0.01, $\mathfrak{s}_{\textit{m}}$=$\mathfrak{s}_{J}$=0.01, $\gamma_{\bm{\textit{x}}1}$=$\gamma_{\bm{\textit{x}}2}$=5000, $\gamma_{\bm{\textit{x}}3}$=1000,  $\gamma_{\bm{\textit{R}}1}$=$\gamma_{\bm{\textit{R}}2}$=1500, $\gamma_{\bm{\textit{R}}3}$=100, $\bm{Q}_1$= $\bm{Q}_2$= $\mathbf{diag}[\frac{1}{20},\frac{1}{20}]$, $\bm{Q}_3$=$\mathbf{diag}[1,1]$.
For model uncertainties setups, in  group A, both the existing and our algorithms were tested  under model-matched plant :  $\bm{J}_0$=$\bm{J}'_0$ and $m_0$=$m'_0$. In groups B and C, the preset reference values  remained unchanged, while the real-plant values increased to $\bm{J}_0$=$\mathbf{diag}[0.688,0.594,0.783](\mathrm{kg}\cdot\mathrm{m}^2)$ and $m_0$=$5(\mathrm{kg})$. 
For disturbances setups, all the comparison groups experienced full disturbances, including $\Delta_{\bm{x}_{i}}$, $\Delta_{\bm{R}_i}$,  $\Delta_{\bm{x}_{i}}^{\parallel}$, $\Delta_{\bm{x}_{i}}^{\bot}$,  $\Delta_{\bm{x}_0}$ and $\Delta_{\bm{R}_0}$. To demonstrate the enhanced disturbance rejection ability, groups B and C respectively experienced extra  strong disturbance forces  $\Delta^{B}_{\bm{x}_0}$  and a moment $\Delta^{C}_{\bm{R}_0}$  exerted on the payload, as given in the caption of Fig.~\ref{Gao4}. 

From the results in Fig.~\ref{Gao4}, the enhanced robustness, including adaptivity to the parametric uncertainties and rejection to the disturbance forces and moment, was validated. For the values of the estimated payload mass $\bm{\bar{m}}_0^{[j]}$ and translational disturbance dynamics $\bm{\bar{\phi}}_{\textit{x}}^{[j]}$  in group B, refer to Fig.~\ref{Gao3_a}, and for the estimated payload inertia tensor $\bm{\bar{J}}_0^{[j]}$ and rotational disturbance dynamics $\bm{\bar{\phi}}_{\textit{R}}^{[j]}$ in group C, see Fig.~\ref{Gao3_b}.

\section{Conclusion}
\label{Conclusion}
In this paper, we proposed an adaptive-neuro geometric control with multiple neural networks and adaptive laws to enhance the robustness of the centralized multi-quadrotor transportation system. Preliminary stability results have been analyzed and the enhanced robustness was demonstrated by numerical simulations. Future work will focus on real-world experiments and a more in-depth stability analysis.
 \vspace{-0em} 

\addtolength{\textheight}{0.5cm}   




\end{document}